# 'I Updated the <ref>': The Evolution of References in the English Wikipedia and the Implications for Altmetrics

Zagovora, Olga*, Ulloa, Roberto, Weller, Katrin, Flöck, Fabian

\* corresponding author: olga.zagovora@gesis.org

With this work, we present a publicly available dataset of the history of all the references (more than 55 million) ever used in the English Wikipedia until June 2019. We have applied a new method for identifying and monitoring references in Wikipedia, so that for each reference we can provide data about associated actions: creation, modifications, deletions, and reinsertions. The high accuracy of this method and the resulting dataset was confirmed via a comprehensive crowdworker labelling campaign. We use the dataset to study the temporal evolution of Wikipedia references as well as users' editing behaviour. We find evidence of a mostly productive and continuous effort to improve the quality of references: (1) there is a persistent increase of reference and document identifiers (DOI, PubMedID, PMC, ISBN, ISSN, ArXiv ID), and (2) most of the reference curation work is done by registered humans (not bots or anonymous editors). We conclude that the evolution of Wikipedia references, including the dynamics of the community processes that tend to them should be leveraged in the design of relevance indexes for altmetrics, and our dataset can be pivotal for such effort.

**Keywords: altmetrics, Wikipedia references, edit histories, data quality, dataset, Wikipedia editors**

## 1. Introduction

### 1.1. Wikipedia references and their challenges for altmetrics

Wikipedia incorporates one of the largest reference repositories in existence. This is primarily due to its guidelines strongly encouraging that all content have to be verifiable, which is mostly achieved by providing a pointer to a reliable source that supports content added to the article text.[1] Thus, Wikipedia articles usually include reference lists; and overall, the English Wikipedia contains more than 55 million references.[2] Cited sources can be different types of publications, including for example formally published scientific papers, books, and news media articles, but also links to websites or any other type of Web documents (Lewoniewski et al., 2017).

These references are exposed to an enormous readership, as Wikipedia is accessed by a wide audience around the world. With more than 300 million page views per day for the English

---

Wikipedia alone, it is one of the top-15 most visited Websites in the world.[3] While recent studies seem to indicate that a large number of users do not fully engage with references by visiting links or retrieving the referenced document otherwise (Piccardi et.al., 2020), references can still make statements more credible simply by appearing alongside them; and they are actively being interacted with more than 32 million times a month - measured by mouse-hovering over the reference footnote (Piccardi et.al., 2020).

Thus, Wikipedia's references have a tremendous impact on its millions of readers - who encounter them while browsing serendipitously or while actively researching a topic. In addition, Wikipedia content, including its references, is incorporated into other data sources and projects, and thus reaches even wider audiences. For instance, Wikipedia content is used as a source for the collaborative knowledge base WikiData[4], which is also used by other platforms. Scholia[5], for instance, creates scholarly profile pages based on WikiData.

Apart from its appeal to the general public, Wikipedia has also become an object of interest in the field of *altmetrics,* an area of research dedicated to studying ways of measuring the impact of scientific work outside of traditional scholarly citation schemes, and often based on social media interactions (Priem et al., 2010; Kousha & Thelwall, 2017). In this context, the value ascribed to Wikipedia as a data source is that it provides an immense repository of literature curated by a large editor community. And with the self-control mechanisms and guidelines applied within this community, Wikipedia references are expected to meet basic quality standards. At the very least they are presumed to be topically relevant and ideally to represent a comprehensive, up-to-date and balanced collection of the most relevant sources. Given the dynamic nature of Wikipedia, it might also be possible to opportunely detect novel and trending publications through the additions and changes to the community-created repository of references. Overall, being cited in a Wikipedia article is considered to be an indicator of some form of impact for a (scientific) publication (Kousha & Thelwall, 2017).

Indeed, Wikipedia data is already considered in altmetrics data implementations (and sold) by aggregators in the field. Currently the most prominent are Altmetric.com[6], PlumX[7], CrossRef[8], and Lagotto[9]. Their indicators are applied in different settings such as publishers' sites or repositories (e.g. institutional or discipline-specific publication databases), and they are used to advertise "impactful" publications. These metrics vary substantially, as aggregators apply different modes of accessing and collecting the data, and there is no standard for detecting or aggregating references on Wikipedia (and several other platforms). Although it can be assumed that data collection is to a considerable degree based on standard document identifiers such as DOIs[10] (Haustein, 2016). However, the specific procedures are not transparent and, thus, altmetrics aggregators have to be viewed as black boxes that could be subject to manipulations (Kousha & Thelwall, 2017), such as researchers adding references to their own publications into Wikipedia articles[11], or even strategic campaigns to insert publications from a specific publisher into Wikipedia articles[12].

---

[3] https://tools.wmflabs.org/siteviews/?sites=en.wikipedia.org, https://www.alexa.com/topsites, as of 15.02.2020
[4] https://www.wikidata.org/
[5] https://tools.wmflabs.org/scholia/
[6] https://www.altmetric.com/explorer/
[7] ttps://plumanalytics.com
[8] https://www.crossref.org/
[9] http://www.lagotto.io/docs/api/

To illustrate some of the challenges in using Wikipedia references as reliable indicators, we will take a closer look at a particular example publication and how it is referenced in English Wikipedia, as identified by our extraction method and dataset (see [Section 3](#)). Our example is based on several references to the publication *"Roy et al. (2001) Structure and function of south-east Australian estuaries. Estuarine, Coastal and Shelf Science 53(3): 351–384."* on different article pages. The first reference to this publication was added to a Wikipedia article in August 2012 ([Figure 1](#), blue line), i.e. more than ten years after the paper's release. Nine months later (1st of June 2013), there were already 53 articles that referenced this publication, all of them done by the same editor (id: 7739861). However, none of the references included its existing Digital Object Identifier (DOI). The corresponding DOI to this publication was added to the existing Wikipedia references during the first quarter of 2014 ([Figure 1](#), orange line), and this was mostly done by one single editor in March 2014 (id: 203434). In November 2018, another editor (id: 15881234) removed 27 instances (50%) of the references although some of them were quickly reinstated.

---

[10]  For example, Altmetric.com is collecting data using the following identifiers https://help.altmetric.com/support/solutions/articles/6000234171-how-outputs-are-tracked-and-measured, and CrossRef is collecting using DOI and landing page URLs https://www.crossref.org/services/event-data/

[11] Wikipedia's guidelines about Conflict of Interest include a section on "Citing yourself", which allows self-citations within certain boundaries. see https://en.wikipedia.org/wiki/Wikipedia:Conflict_of_interest. To the best of our knowledge, there are no studies that investigate in detail how common self-citations are in Wikipedia or that aim to identify misconduct in the area of self-promoting scientific articles through Wikipedia.

[12]One example can be found at: https://web.archive.org/web/20200323131800/https://annualreviewsnews.org/2020/02/25/seeking-a-wikipedian-in-residence/

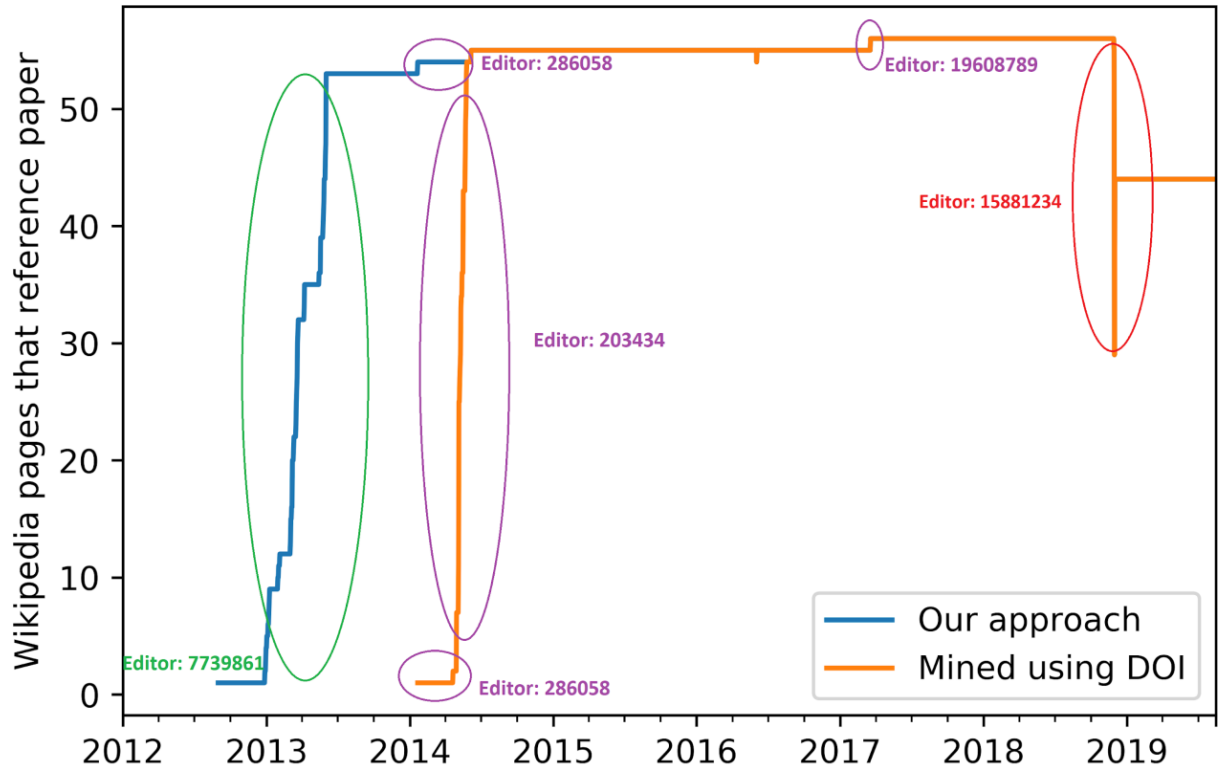

**Figure 1 - References in the English Wikipedia for one example paper, as identified by our approach (blue line) and by approaches based only on document identifiers (orange line).** Areas highlighted by circles correspond to edits made by one specific Wikipedia user (see editor ID): the green circle indicates an editor adding instances of the reference without any document identifier, violet circles represent different editors who modified existing references (e.g., by adding a DOI) and red for editors who deleted references from articles.

Despite the widespread popularity and importance of Wikipedia and the practical role it plays in applied altmetrics, this basic example illustrates several issues that motivated our work. First, it highlights a weakness of mining references based only on document identifiers (orange line), that lead us to create an alternative method that uses the entire text of the reference (blue line); the former misses the reference for the first two years of its existence. Second, it shows the impact that a single editor can have on the visibility of a reference by systematically adding or removing it from different articles. Third, it exposes the lack of understanding about Wikipedia editors as the creators and curators of Wikipedia and their impact of references being implemented. Together with the intransparency of aggregated altmetrics indicators, it exposes a current gap in our understanding of the nature and quality of Wikipedia references in altmetrics.

At the same time, the example captures the value of our investigation as an important step to close this gap. It suggests that the *origin and evolution* of references inside Wikipedia articles can disclose anomalies in the activity around Wikipedia references and that many of the collaborative negotiation processes that govern the inclusion, modification, and deletion of references are encoded within article revisions, and therefore can reveal information about the editor community responsible for the maintenance of this valuable asset.

## 1.2 Research focus and contributions

With this in mind, we outline the specific research areas which our research questions and contributions fall into.

**Insights into reference evolution over time.** We are interested in the development of references in Wikipedia from a longitudinal perspective. The ongoing transformation and expansion of Wikipedia content affects the potential (measured) impact of cited sources by constantly increasing or decreasing the number of instances that reference them, either by internal changes in the articles or by introduction and deletion of entirely new articles. Therefore, Wikipedia presents a scenario that is very different from other settings in citation analysis in areas of bibliometrics and altmetrics. In bibliometrics based on citations from traditional publication outlets, both (1) alterations to existing references lists and (2) retractions of entire articles are rare events (Shema et al., 2019). Traditional citations typically produce *growing* numbers that are then incorporated into different bibliometric indicators (citations add up as more publications that cite a source are published and almost none of them disappear again).

In contrast to bibliometrics, the altmetrics field has to deal with fluid types of data sources, as they include dynamic material[13] such as tweets or Facebook posts that might be deleted or modified. Wikipedia adds another challenge. While many other altmetric sources are based on *collective* approaches in which individuals act on their own and indicators merely cumulatively aggregate these individual activities, Wikipedia as a *collaborative* system relies on a consensus between members that can take time to reach an equilibrium, and which might be perturbed again as new information becomes available. References may be added by one person, removed by another, and then re-inserted or edited again. These processes can happen multiple times, and little is known about how this has affected Wikipedia's references in the past and how many editing activities are performed on references overall.

---

[13] While in fact, social media content containing altmetrics indicators (e.g., Facebook posts) is deleted to some extent after the initial altmetrics detection, we are not aware of aggregators' metrics that take these deletions into account. To the best of our knowledge, most aggregators are removing only deleted Tweets as per terms of Twitter data usage.

This leads to our first research question: (*RQ1) How do Wikipedia references evolve over time?* We examine the fluctuation of all references of Wikipedia by analyzing the number of actions performed on them. To the best of our knowledge, we provide the first longitudinal study of the evolution of references across all revisions in the English Wikipedia by also considering all references - not only those that include a standard document identifier (DID), such as the Digital Object Identifier (DOI).

For practical reasons, altmetrics indicators often rely on DIDs for the detection of publications, leading to the question: (RQ2) *What is the current and past coverage of references that include DIDs?* The references that lack DIDs are simply missed by methods that rely solely on them, and their visibility is therefore decreased. We will tackle this question by estimating, at different points in time, the proportion of references that include DIDs, and by using current knowledge from our 2019 dataset to calculate which references lacked DIDs in the past.

***Insights into the editors of Wikipedia references.*** We are interested in getting a better understanding of *who* adds, modifies, or deletes Wikipedia references. Currently learning more about the people who actually produce the social media contents that are then considered sources for altmetrics is in its beginnings (Holmberg, 2015; Imran et al., 2018). For example, little is known about who actively tweets, blogs, and posts about science - or incorporates references as sources for encyclopedic articles.

We, therefore, set out to answer *RQ3: Who creates and maintains Wikipedia references, and in which way?* This question pertains, on one hand, to which parts of the Wikipedia editor base engage in different reference-related activities, e.g., occasional users, very active registered editors, unregistered users, or automated bots. On the other hand, discovering patterns of interaction with references exhibited by editors that tend to references, such as focussing their actions mostly on reference maintenance as opposed to only adding references occasionally as part of other writing efforts. This more fine-grained picture of possible roles of editors in the reference ecosystem can help to understand the editor community that is responsible for the activity around the Wikipedia references.

***A comprehensive reference dataset based on edit histories.*** Addressing these and future research questions becomes only viable with a novel dataset (Zagovora et al., 2020) of individual revision histories of all Wikipedia references[14] ever created in the English Wikipedia until June 2019 together with information about editorship. We created the dataset by leveraging WikiWho[15] (Flöck & Acosta, 2014), a service that tracks the additions, changes, and reinsertions of words (tokens) written in Wikipedia (see Section 3 for details). With our new method, we are able to track the types of actions performed on references as well as the editors that contribute and maintain references in the English Wikipedia. We evaluated our dataset with crowdworkers and demonstrated its high accuracy, despite our method not relying on any types of document identifiers for tracking references to reach the maximum coverage.

---

[14] "reference" defined as the content included inside a Wikipedia <ref> tag
[15] https://www.wikiwho.net/

We find that the quality of Wikipedia references is in continuous improvement based on constant activity, including focussed efforts to add DIDs, and the fact that registered humans (as opposed to bots or anonymous editors) are mainly responsible for the curation of the references. Nonetheless, we believe that the reliance on DIDs is not currently sufficient to capture all relevant publications cited in Wikipedia, and the specialization of editors in certain actions (e.g. only adding or deleting references) deserves more attention to discard the possibility of exploitative behaviour.

Notwithstanding the pitfalls of an open, collaborative system where little control exists over the content, the historical record of Wikipedia can be used to improve methods of mining and ranking the relevance of references. We recommend that altmetrics indicators should leverage the Wikipedia revisions to decrease manipulations, increase coverage, and assign impact based on past activity and the editor community that surrounds each reference. Our dataset could be pivotal in developing such improvements.

The rest of this paper is organized as follows: Section 2 will offer an overview of the related work relevant for Wikipedia references and altmetrics, Section 3 is dedicated to the description of methods to build the dataset, Section 4 presents a gold standard dataset that is used for the evaluation of our references dataset, Section 5 presents general statistics of the Wikipedia references and main findings regarding our research questions, and Section 6 will conclude and summarize our findings.

## 2. Related work

The most comparable dataset to the one we provide is presented by Halfaker et al. (2019) and Redi & Taraborelli (2018). They also include some form of *historical* data about references in Wikipedia. However, their work differs from our approach as the authors relied on the presence of standardized DIDs as part of the reference and thus were (1) not capturing all references, and were (2) assigning editors and timestamps of origin to references according to the Wikipedia revision in which the identifier was included, even if in fact the reference as such was created earlier (cf. Figure 1). Lastly, (3) modifications and deletions done to the references after the inclusion of the identifiers were not tracked. While the dataset has been publicly shared with the interested community and was used, e.g., to study topics of citations, to the best of our knowledge no-one has yet used it to study the evolution of references or editing behaviour related to references.

Other works only provide *static* (non-historical) snapshots of references in Wikipedia language editions, like Nielsen (2008)[16] or Singh et al. (2020), that were created for specific tasks. Nielsen (2008) used the "cite journal" template from references to create a dataset of journal papers that were cited in Wikipedia pages. This dataset was then used to cluster Wikipedia pages and corresponding scientific journals into distinct research topics. Singh et al. (2020) created a dataset of references and classified them into 3 groups: journal articles, books, and other Web content.

Recently, research has started to look more closely at how Wikipedia *readers* interact with references. With Wikipedia references being actionable items that users can click on, they have been described as a "bridge to the next layer of academic resources" (Grathwohl, 2011). However, recent studies (Redi, 2018; Piccardi et al., 2020) show that not all references are being equally visited by Wikipedia readers. Piccardi et al. (2020) conclude that regarding references "readers are more likely to use Wikipedia as a gateway on topics where Wikipedia is still wanting and where articles are of low quality and not sufficiently informative". They found that in the large majority of cases where Wikipedia articles are of high-quality readers do not make use of the references but stay at the Wikipedia article as the "final destination" of their information journey (Piccardi et al., 2020). This kind of work gives us more insights on the consumer perspective of Wikipedia references, which adds to the general perspective of how Wikipedia is used, e.g. how Wikipedia articles are read or how people are citing from Wikipedia articles (Bould et al., 2014; Okoli et al., 2014).

To the best of our knowledge, there are only few studies focusing on editors as the creators of references in Wikipedia and thus contributing to the *producer* perspective. With a comparatively small data sample (~5000 articles), Chen & Roth (2012) showed that "a reference occurs when a set of committed and qualified editors are attracted to the article". Huvila (2010) conducted a survey of Wikipedia editors, also including questions broadly related to reference editing. Specifically, the survey enabled them to differentiate editors based on their information behaviour and the sources the editors were using for editing articles. The results indicate a preference for sources that are available online. There is also some specific, ongoing research on other and more general perspectives on the producer side of Wikipedia, e.g. on who edits Wikipedia, general editing patterns (Flöck et al. 2017), who becomes a power editor (Panciera et al., 2009), or how editors collaborate (Kittur et al. 2007; Murić et al, 2019).

---

[16] The dataset is available via  http://hendrix.imm.dtu.dk/services/wikipedia/citejournalminer.html

Furthermore, in the area of *altmetrics* research, a certain focus has been placed on untangling the relations between references in Wikipedia and the scientific publications they are referring to. For example, altmetrics studies have scrutinized the relevance of scientific publications mentioned on Wikipedia (Sugimoto et al., 2017; Kousha & Thelwall, 2017). Shuai et al. (2013) found that papers, authors, and topics that were used as references on Wikipedia have higher citation counts than those that were not mentioned. At the same time, only a narrow part of influential works is cited on Wikipedia (Kousha & Thelwall, 2017). Nielsen (2007) showed that citations from Wikipedia are correlated with the total number of journal citations, whereas the correlation was weak with the journal impact factor. Yet, according to Nielsen (2007), Wikipedia editors tend to cite articles from high impact journals such as Nature, Science, or New England Journal of Medicine. Teplitskiy et al. (2017) conducted a similar experiment with a newer dataset and found that not only impact factor increases the probability of a paper being mentioned on Wikipedia, but also open access principles. According to Mesgari et al. (2015), the *quality of content* and of referenced sources, in particular, was one of the major study objects on Wikipedia. For example, Lewoniewski et al. (2017) studied the similarity of sources from different Wikipedia language editions. They found that URLs in references shared many domain names between language versions, but there were not many cases of exact matches of URLs in references across languages. Lin & Fenner, (2014) showed that ecology and evolution are better covered with references from PLOS than other subjects. Nevertheless, these results might not show the full picture while references were reported as incomplete and accompanied by the lack of standardization (Pooladian & Boorego, 2017).

The altmetrics community is interested in, for example, learning more about whether being cited in Wikipedia articles indicates that a scientific publication has an impact beyond academia into the broader public (Lin & Fenner, 2013; Thelwall, 2016). Lin & Fenner (2013) argue that Wikipedia references might capture a "discussion" group, one of the engagement types with research publications. We provide some first insights in this description together with descriptive statistics that we hope will inspire more detailed reflections on how to interpret different types of reference editing behaviour patterns. This will have to be translated into a broader discussion about how the altmetrics community wants to define the *impact* of a Wikipedia reference (considering its edit history), e.g. when reporting Wikipedia citation counts for a specific publication. For example, our dataset can enable a finer analysis of the revisions of references that can help to detect potential disruptions (e.g. sudden appearance of the same reference across various articles, or highly active individual editors who are responsible for large numbers of new references).

Our work may thus contribute to the theoretical value of being cited by a Wikipedia article related to another area of altmetrics research focused on understanding the *quality of data* obtained from aggregators. For example, Zahedi & Costas (2018) and Ortega (2018) have started to compare different altmetrics aggregators to illustrate potential challenges for data quality. Differences start with coverage by aggregators. In the context of Wikipedia, this means that references appearing on Wikipedia make up from 2% of publications tracked by Altmetric.com up to 5.1% of those tracked by Lagotto. Those differences are due to the aggregator's methodology and the datasets of publications they are tracking (Zahedi & Costas, 2018). These studies also observe different mean values for how often publications are mentioned on Wikipedia: publications in the Altmetric.com collection are on average included by 1.7 Wikipedia pages, publications in the Lagotto collection are on average cited by 2.9 Wikipedia pages, and publications in CrossRef Event Data are on average cited by 15.7 Wikipedia pages (Zahedi & Costas, 2018). We assume that these wide differences are not only due to the diverse sets of publications covered by the aggregators but also due to their distinct methods of tracing Wikipedia references that are prone to various errors considering the challenges inherent to Wikipedia data. Besides the difficulties of keeping track of continuous changes in Wikipedia where references may be modified or removed, one important error source is the reliance on standard document identifiers to trace publications (Ortega, 2018). Nevertheless, even other approaches that rely on title and first author name (Kousha & Thelwall, 2017) may fail to extract mentions from incomplete references (Pooladian & Borrego, 2017). Given the quality of our dataset, it has the potential to serve as an external base for comparing different data collection approaches used by altmetrics aggregators, giving them the opportunity to increase their coverage and impact indexes by looking at different points in time of the revision history.

## 3. Creating the reference histories dataset

We will first introduce our dataset that underlies the following analyses. The dataset is based on the revisions of all articles in the English Wikipedia edition since its origin until June 2019 and contains the change history of all individual references. References, i.e., pointers to external sources (which may be any type of document, including academic and non-academic publications), are inserted into Wikipedia in a standardized way. They appear as "inline citations"[17] in the main body of the article, formatted by `<ref>` … `</ref>` markers in Wiki markup. Based on these ref tags, they can be identified inside the main text and Wikipedia also uses them to create the reference lists at the end of the article. For our work we consider all inline citations marked by ref tags.[18] Typical examples of reference formats would be the following:

---

[17] The Wikipedia community utilizes the term "inline citation", which broadly speaking corresponds to the "in-text citation" as known from bibliometrics. See more details here https://en.wikipedia.org/wiki/Wikipedia:Inline_citation
[18] References are generally created with Wiki markup language, by adding <ref> tags around the source. Additionally, some references can be added automatically by dedicated templates. We are not considering materials that are not referenced as inline citations (e.g., publications from the "Additional reading" section) as the guidelines recommend to include references via <ref> tags (inline citations) as the standard (https://en.wikipedia.org/wiki/Wikipedia:Citing_sources).

- `<ref>Clark 1971, p.18</ref>`
- `<ref>Clark, Ronald W. (1971). Einstein: The Life and Times. ISBN 0-380-<44123-3</ref>`
- `<ref name=ronald>{{cite book |last=Clark |first=Ronald |title=Einstein: The Life and Times |publisher= |date=1971 |page=18 |url= }}</ref>`.

In the following subsections, we (3.1) highlight the idiosyncrasies of Wikipedia and its references and the key challenges in tracing the revision histories of individual references from articles revision histories. We then (3.2) describe our solution in detail, and (3.3) explain how we extract document identifiers (DIDs) for the references.

## 3.1 Wikipedia Articles: Revisions, References and Tokens

The main content corpus of the Wikipedia encyclopedia is organized in articles. Each article $A$ consists of an ordered list of revisions $R$, i.e. $A = [R_0, \ldots, R_n]$, where each revision is a new version of the text which was contributed by editor $e$ at timestamp $z$. Inside the Wiki markup text, references are added inline, immediately after the facts they support and are displayed as footnotes at the bottom of the Wikipedia article in a dedicated section. They are bracketed by `<ref>` … `</ref>` tags. For the front-end HTML representation, the ref tags are converted by a Wikitext parser into footnote references with an appropriate order.

The revision history of a reference is given by the article revisions in which it was added, or edited, either in its entirety or partially. Edits can be performed by registered Wikipedia user accounts (including automated scripts/bots), or through non-registered sessions represented via an – mostly dynamic – IP address (see [Subsection 5.2](#) for a more detailed typology of editors). As each revision within an article is associated with exactly one editor $e$, so is each creation or change action performed on a specific reference through that revision.

The revision history of a reference within a specific Wikipedia article starts with the reference's first creation. After a reference is *created*, it can be *modified*, e.g., by correcting the name of an author. A reference can also be *deleted*, and then be *reinserted* in its entirety after deletion. Generally, a considerable part of editing activity in article revisions concerns the deletion and reinsertion of entire content sequences, for example in edit disputes (Flöck et al. 2017).

Identifying the specific revisions in which the above-described changes are applied to a given uniquely identified reference in Wikipedia presents two major challenges.

1. Tracking changes of any particular target sequence in a long text, and specifically Wikipedia articles, can be error-prone if either multiple other changes have been applied in the same revision and/or the sequence in question was moved to a remote part of the document, which happens regularly in Wikipedia (Flöck & Acosta, 2014). In these instances, common text difference algorithms (such as employed by standard Mediawiki instances or text mining tools) can lose track of sequences and erroneously assign them as new content or as deleted.[19]

2. Even if all concrete tokens (i.e., strings that are either words or combinations of alphanumeric characters) constituting a reference are correctly tracked, deciding if a reference is identical to another reference in two consecutive article revisions is non-trivial, if either a majority of the tokens have been replaced or altered or if key tokens have been changed (e.g., the content of the "title" or "year" fields). This might indicate either a small correction to an existing reference or, in this example, its replacement with a new edition of the underlying publication, which is not identical from a bibliometric point of view.

For the second issue, text similarity measurement (e.g., cosine similarity) based on strings of tokens might be an apparent choice for a potential solution. However, within the limited scope of a Wikipedia article, its references could be very similar in their text strings due to the recurrence of technical terms that are germane to the article's topic. The intrinsic structure of references, that follows standardized formats may further accentuate the perceived and measured similarity of text strings, making it likely to have two similar but distinct references being considered as the same one. For example, the reference '*Darwin (1859). On the origin of species*' has less formal similarity to (i) '*Darwin, Charles (1859). On the origin of species by means of natural selection. London: John Murray*' than to (ii) '*Crawford (1859). (Review of) On the origin of species*' - and yet (i) corresponds to the same reference, but (ii) does not.

To address these issues, we take advantage of WikiWho, an approach that solves the change attribution problem at a token level with over 95% accuracy (Flöck & Acosta, 2014). Each token ever inserted in an article has been assigned a token ID that uniquely identifies it through all revisions. WikiWho uses the relative location of tokens in sentences and paragraphs to accurately track each of them between every pair of revisions. Figure 2 illustrates the allocation of WikiWho token IDs for the two first revisions of an article.

---

[19] Cf. Wikipedia diffs (https://en.wikipedia.org/wiki/Help:Diff) and Flöck & Acosta (2014) for a more general discussion.

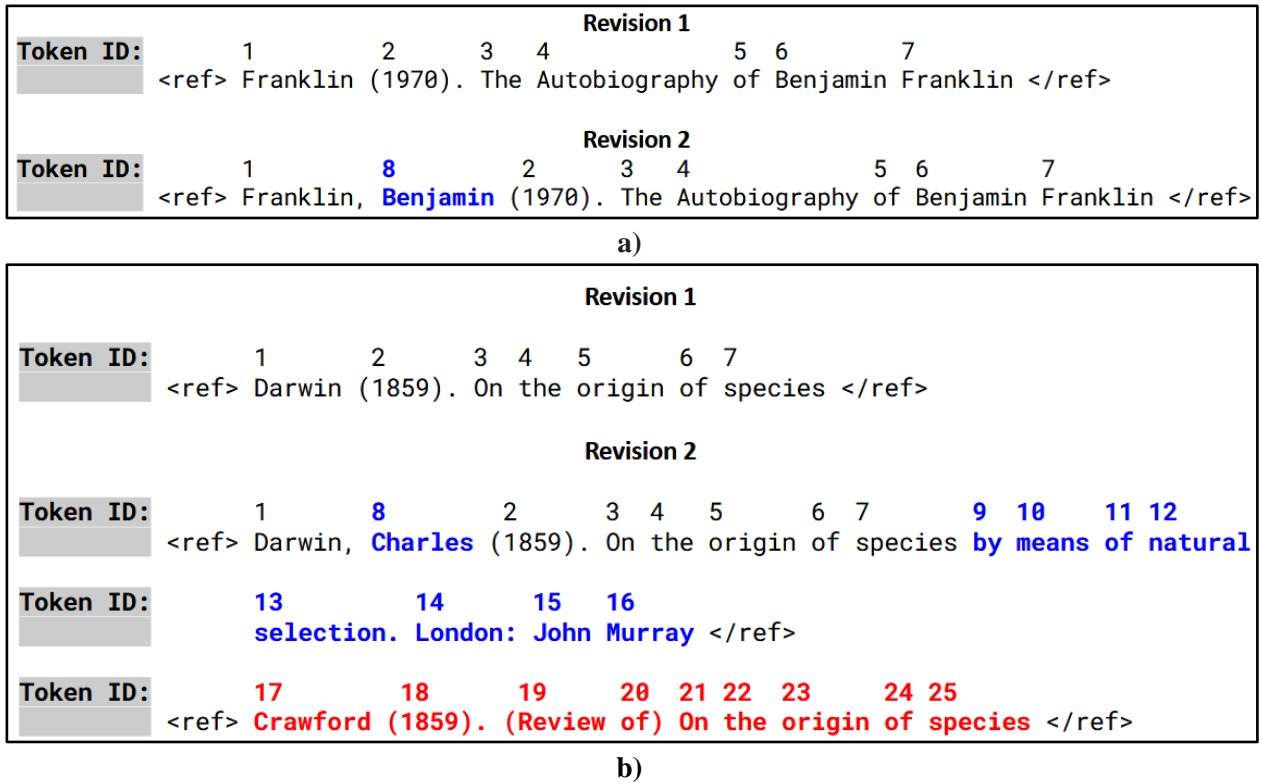

**Figure 2 - Examples of token ID assignments by WikiWho before and after an edit. a)** In Revision 1, the different tokens are identified from 1 to 7. In Revision 2, *"Benjamin"* (blue) is inserted and WikiWho assigns a token ID (8). Note how the older instance of *"Benjamin"* (ID 6) is tracked as a distinct token, merely sharing an identical string. **b)** In Revision 1, the different tokens are identified from 1 to 7. In Revision 2, *"Charles"* and *"by means of ..."* (blue) are inserted and new token IDs (8-24) are assigned. Another new reference "*Crawford (1859)...*" is added in Revision 2, in a paragraph further below (position not shown here). Note that new token IDs are assigned despite shared strings with the previous reference. This is achieved as WikiWho takes the larger context of a changed sequence into account. These toy examples do not track punctuation for simplicity, while WikiWho does so in practice.

As references are nested inside XML tags inside sentences inside paragraphs, WikiWho is particularly well suited to uniquely identify tokens of references. This solves our first issue of change tracking. Secondly, instead of using a string similarity measure, we proceed with the results of WikiWho and apply Jaccard similarity between the token IDs of references to match them between revisions, drastically reducing the false-positive cases. We describe this approach in detail in the next Subsection and the evaluation in Section 4.

## 3.2 Extraction of reference histories

Our dataset of references is organized per Wikipedia article. I.e., we do not – for this work – match references across articles. Formally, for each article A, the dataset contains a list of tuples $H_f = [< a_{f_0}, t_{f_0}, r_{f_0}, h_{f_0}, e_{f_0}, z_{f_0} >, \ldots, < a_{f_n}, t_{f_n}, r_{f_n}, h_{f_n}, e_{f_n}, z_{f_n} >]$ that represents the history of actions $a_{f_i}$ ('creation', 'insertion', 'deletion' or 'reinsertion') performed over reference $f$, where:

- $t_{f_i}$ is the list of WikiWho token IDs that were part of the reference in revision $r_{f_i}$
- $h_{f_i}$ is a hash value calculated of over $t_{f_i}$,
- $e_{f_i}$ is an editor that performed an action $a_{f_i}$ at time $z_{f_i}$, and
- $H_f$ is sorted according to time $z_f$, $z_{f_0}$ is the oldest reference.

To build this dataset, we first mine all inline citations of all Wikipedia revisions using the WikiWho token IDs that correspond to the string tags `<ref>...</ref>` and `<ref name=…>...</ref>`; the void tags, i.e. the one-sided tags (`<ref name=... />`) are excluded because they correspond to duplications of existent references. For each revision $R_i$ in each article $A$, we then have a list of references that belong to that revision $G_i = [f_0, \ldots, f_m]$ where each reference $f_j$ is a tuple $< t_j, h_j, e_j, z_j >$.

The next step is to associate the references in $G_i$ to those in $G_{i+k}$, so that two references are added to $H_f$ if they are *equivalent*, i.e., referring to the same publication.

In trivial cases, a reference $f$ does not change between article revisions so we use the hash values to match all identical references across all G, and we store the matched references of $f$ in $H_f$. For now, each $H_f$ is incomplete as there could be two references histories $H_f$ and $H_g$ that belong together because with this procedure even a small modification is enough to change the hash value. Therefore, all actions $a_f$ are tagged as 'unknown'.

In the nontrivial cases, the references have been modified between two consecutive revisions. We then rely on a combination of Jaccard similarity between the lists of WikiWho token IDs of references in H and G. The following procedure was applied for each reference $f$:

1. (**Creation**) Select the oldest tuple $< r_{f_o}, h_{f_o}, a_{f_o}, e_{f_o}, t_{f_o}, z_{f_o} >$ in $H_f$. Search $R_j$ in $A$ that corresponds to $r_{f_o}$ and switch the value of $a_{f_0}$ from 'unknown' to 'creation'. **Go to Step 2.**
2. (**No action**) If the successor revision of $r_{f_i}$ is also in $H_f$, i.e. $r_{f_i+1} = R_{j+1}$, then:

1. If hash $h_{f_i}$ of reference fi is the same as hash $h_{f_{i+1}}$, i.e., $h_{f_i} = h_{f_{i+1}}$, then remove tuple $< r_{f_{i+1}}, h_{f_{i+1}}, a_{f_{i+1}}, e_{f_{i+1}}, t_{f_{i+1}}, z_{f_i} >$ (no action occurred between $r_{f_i}$ and $r_{f_{i+1}}$, i.e. they have the same hash).
2. Otherwise, switch the value of $a_{f_{i+1}}$ from 'unknown' to 'modified'.
3. **Go to Step 2** using the next oldest tuple $< r_{f_{i+1}}, h_{f_{i+1}}, a_{f_{i+1}}, e_{f_{i+1}}, t_{f_{i+1}}, z_{f_{i+1}} >$ in $H_f$.

3. **(Modification)** Otherwise, we identify a list of candidate references in $R_{j+1}$, i.e. reference hashes that did not exist in revisions before $R_{j+1}$. A reference is a candidate c if $r_{c_0} = R_{j+1}$, where $< r_{c_{i+1}}, h_{c_{i+1}}, a_{c_{i+1}}, e_{c_{i+1}}, t_{c_{i+1}}, z_{c_{i+1}} >$ in $H_c$.

    1. For each candidate reference c, we calculate the Jaccard similarity between its tokens $t_c$ and $t_{f_i}$.
    2. We select the candidate with the highest Jaccard similarity, given that it is higher than 0.2 (see Supplementary materials for the threshold selection process).
    3. If all candidates have a similarity lower than 0.2, then we check if there is a candidate in which $t_c \subset t_{f_i}$, i.e. all tokens of the previous reference are reused. This condition targets short references that are extended by many tokens in the new revision leading to very low Jaccard similarity.
    4. If a candidate is selected, then add the tuple $< R_{j+1}, h_c, 'modified', e_c, t_c, z_c >$ to $H_f$. Then merge $H_f$ and $H_c$ and **go to Step 2** using that tuple.
    5. Otherwise, if no candidate is selected, **go to Step 4.**

4. **(Deletion)** Insert the tuple $< r_{f_{i+1}}, h_{f_i}, 'deleted', e_{f_i}, [], z_{f_i} >$ to $H_f$ and **go to Step 5**.

5. **(Reinsertion)** Move to the next revision in $A$, i.e. let $R_j$ be $R_{j+1}$.

    1. If there exists $r_{f_i}$ in $H_f$, such that $r_{f_i} = R_{j+1}$; then switch the value of $a_{f_i}$ from 'unknown' to 'reinserted' and **go to Step 2** using $< r_{f_i}, h_{f_i}, a_{a_i}, e_{f_i}, t_{f_i}, z_{f_i} >$ in $H_f$.
    2. Otherwise, use steps 3.1, 3.2, and 3.3, in order to select a candidate in $R_{j+1}$ that matches $t_{f_i}$.
    3. If a candidate c is selected, then add the tuple $< R_{j+1}, h_c, 'reinserted', e_c, t_c, z_c >$ to $H_c$, and merge $H_f$ with $H_c$ and **go to Step 2** using that tuple.
    4. Otherwise, if no candidate is selected (the reference was not yet reinserted). **Go to Step 5**.

After this procedure, we obtain four types of actions for each reference:

(1) **creation**, the singular revision in which the reference appears for the first time,

(2) **modifications**, revisions in which the text of the reference was changed, e.g., adding full names of the authors or introducing a DOI; here we explicitly mean that the cited source of the reference remains the same despite changes in reference presentation,

(3) **deletions**, revisions in which the reference was removed from the Wikipedia article,

(4) **reinsertions**, revisions in which the same reference was added again after being removed.

### 3.3 Tracking of DID references

The content of a reference can also contain different types of document identifiers (DID) that have been assigned to the referenced source during its publication process, e.g., Document Object Identifier (DOI). DIDs can easily be used to unambiguously trace individual references, both within Wikipedia and outside of it. They are also often used to trace references in different contexts in the area of altmetrics. In our approach, we are able to extract and monitor all references in a Wikipedia article. However, we take a closer look at the subset of references containing DIDs for two reasons: Firstly, this enables comparisons with previous works which have relied exclusively on document identifiers to extract references for Wikipedia articles. Secondly, Wikipedia includes references to publications that range from strictly refereed and well-reputed scientific outlets to everyday blogs, twitter profiles, and Reddit posts, and we aim to narrow the focus of our investigation to such publications relevant to altmetrics and the academic community. Although DIDs can be an indicator that a reference is academic[20], we are mindful that references with DIDs are not necessarily academic works. Yet, they provide a viable filter to concentrate on references relevant in the context of this work.

As one aspect of the evolution of Wikipedia references over time, we look at when DIDs are added to references in the version history. This allows us to estimate how many references are missed by approaches that rely solely on the presence of DIDs for identifying and counting Wikipedia references. The missed references comprise not only those that by their nature do not have an identifier (e.g. a news article) but also those that did not include a DID at the time of the reference extraction although they have been assigned a DID outside of Wikipedia that could be later on inserted.

We distinguish between several types of references based on DID information ([Table 1](#)). The term *DID-Reference* (DID-R) corresponds to references that by the time of our data collection (June 2019) had a DID. If the DID was immediately included when the reference was created, we refer to it as *DID-Born Reference* (DBorn). Otherwise, if the DID was added after the reference was created, we call it *DID-Lagged Reference* (DLag). Their counterparts, i.e. references that by data collection date did not have a DID, are called *No-DID References*. Note that this classification depends on the time of data collection, as some of the DID-Lagged References would have been classified as No-DID References in previous years and current No-DID References may still receive a DID at a future point in time.

---

[20] We use the term "academic" instead of "scientific" to indicate the inclusion of all works not only from "harder" sciences but also from social sciences and humanities. This is in line with Halfaker et al. (2019).

**Table 1. Types of references according to if and when a DID was added.** The first and second column indicate the names that we use to identify the type and subtype of reference respectively. The third column describes the subtype of references based on when the DID was added.

| Type | Subtype | Description |
|---|---|---|
| **DID Reference (DID-R)** | **DID-Born Reference (DBorn)** | References that already included a DID when they were created. |
| | **DID-Lagged Reference (DLag)** | References that did not include a DID when they were created, but were assigned a DID at a later point, before the time of our data collection. |
| **No-DID Reference (No-DID)** | | References that did not include a DID by the time of data collection. These **might** receive a DID later (after our data collection), if a DID, in fact, exists for the referenced publication. |

After we rebuild the history of all references for each Wikipedia article as explained in the previous subsection, we proceed to extract the DIDs on all the versions of each reference. We used modified versions of regular expressions based on Halfaker et al. (2019) to extract the following DIDs: Document Object Identifier (DOI), International Standard Book Number (ISBN), PubMed Identifier (PMID), PubMed Central identifier (PMCID), International Standard Serial Number (ISSN) and arxiv.org Identifiers (ArXiv ID). Once we extract the DIDs, we can retroactively recognize the DLag references and their content ($t_{f_i}$) as our dataset already contains historical information of each reference ($H_f$). Our method properly handles cases in which a reference has two identifiers (e.g. correction of a DID, or one DOI and one ISBN). We keep the timestamp ($z_{f_i}$) and editor ($e_{f_i}$) that introduced or modified the DID, so that we can further analyze the dynamics of creation and addition of the DIDs.

# 4. Evaluation of the reference change tracking method

In this section, we evaluate the performance of our method for tracking version histories for references. First, we describe a gold standard dataset that we created for evaluation purposes using crowdworkers. Then, we present the overall performance. Last, we compare our method to a baseline relying on cosine similarity.

## 4.1 Gold Standard Dataset

To make sure that our method correctly identifies references in different forms across histories, we created a gold standard dataset of 952 pairs of references, each pair similar to the example in Figure 2a. The pairs are labelled as *Equivalent* or *Distinct,* depending on whether each pair corresponds to the same bibliographical resource or not. Each pair of references were judged by at least three

FigureEight[21] crowdworkers. Each worker indicated if the pair corresponds to the (1) same resource, (2) different resources or (3) if it was not clear. See Appendix A for the instructions we provided for FigureEight crowdworkers, an example question, and a note on fair payment (Zaldivar et al., 2018).

If the agreement[22] between the workers falls below the limit of 0.7, additional persons are assigned until the agreement reaches the required limit (0.7), or until at least five workers have made judgments. Prior to the task, each worker was trained using 115 examples that illustrated different cases, and they had to correctly label at least 5 out of 6 test pairs of references. Moreover, all the answers from a given worker were removed (and a new worker assigned) if their accuracy fell below 0.8. Training and test items have been pre-labelled by the authors of this paper.

In total, 1000 items were presented to the workers, out of which 952 were labelled as either Equivalent or Distinct. We were not able to classify 48 pairs of references because five assigned workers did not agree above the 0.7 limit. One of the researchers closely inspected these cases[23] and confirmed that the low agreement score stemmed from the ambiguity of the items. For example, it is not possible to decide between the pair *<http://www400.sos.louisiana.gov:8090/cgibin/?rqstyp=elcmpct&rqsdta=1021952051300605, http://www400.sos.louisiana.gov:8090/cgibin/?rqstyp=elcpr&rqsdta=10050205>* as the references contain the same domain but the value of the last URL parameter ("*rqsdta*") is different, visiting the URL does not help as the page does no longer exist.

Cases like the above were included in the workers' task as we did not want to distort the true state of Wikipedia references. The set of 1000 items was taken using a stratified random sample from all the references in Wikipedia revisions (Appendix B). The set consists of 8 strata with similarities from 0 to 1 with 0.125 steps, and 125 pairs of references per stratum. Therefore, we make sure that our sample is representative of the Jaccard similarity scale we used in our method; given the accuracy of WikiWho, we knew a priori that most pairs of references fall into the extreme values of similarity (i.e. 0 or 1).

---

[21] www.figure-eight.com

[22] We adopted the "confidence score of the row" of the Figure Eight platform. This value describes the level of agreement between multiple contributors, where the sum of the contributors' trust scores of the most common answer is divided by the sum of the trust scores of respondents to that question. See details here https://success.appen.com/hc/en-us/articles/201855939-How-to-Calculate-a-Confidence-Score

[23] The inspection was done using contextual information from the text surrounding the references in previous revisions, testing URLs, and using external resources (e.g. search engines, archive.org).

## 4.2 Performance

We compared the 952 pairs of references labelled by the crowdworkers, against the labels assigned using our method. Figure 3 illustrates the performance metrics for different Jaccard similarity thresholds in our method (see Step 3.3 of Subsection 3.2). Based on this data, we selected a threshold of 0.2, a good trade-off between precision and recall. At that threshold, the method equally balances the labelling errors between false positives and false negatives.

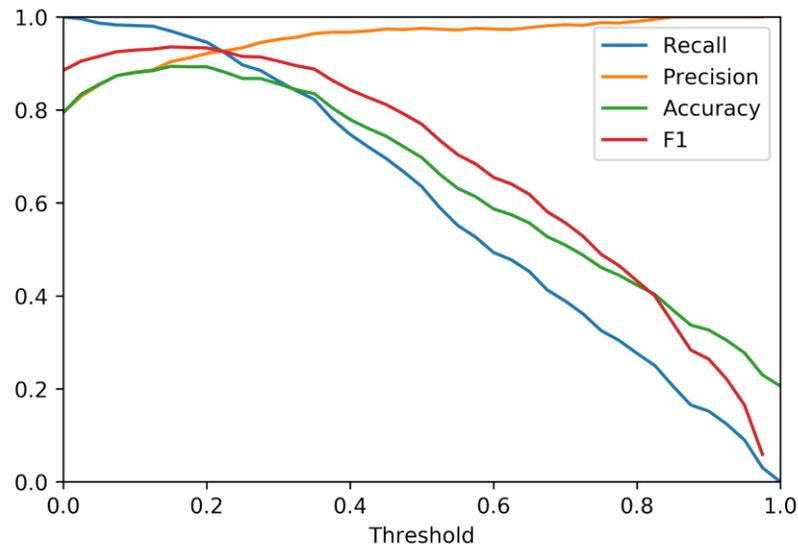

**Figure 3 - Performance metrics for identifying equivalent references.** X-axis shows the threshold of Jaccard similarity between Wikiwho token IDs and Y-axis shows the Precision (Blue), Recall (Orange), Accuracy (Green), and F1 (Red) scores.

To find the overall performance metrics for our method we resampled our stratified sample so that it is representative of the original distribution of Jaccard similarities (calculated with a 100,000 sample) of pairs of references extracted in the same fashion as described in Subsection 3.2. Table 2 presents the micro-average performance metrics for the identification of the same and different references between revisions.

**Table 2 - Micro-average performance metrics for the labelling of pairs of references.** The three metrics are calculated so that they represent the original distribution of Jaccard similarities in the method by resampling from the stratified sample. The metrics show the micro-average performance, so these are the expected overall score, in which each evaluated pair of references contributes equally to the score (regardless of the strata they belong to).

| Precision | Recall | F1-score |
|---|---|---|
| 0.96 | 0.96 | 0.96 |

Upon careful examination of the 48 cases (in which the crowdworkers could not decide) we found that in 30 cases our method is able to appropriately decide based on the contextual information that is encoded in the WikiWho data model. Without this context, some reference pairs might be perceived as equivalente. For example, there is an article about a music artist that has several references to different music albums from the same platform. By quick inspection of the URL without additional context, one might think that it is an equivalent reference. In this case, our approach would differentiate these references if they have been placed at different (relative) positions in the article, despite high string similarity.

## 4.3 Baseline Comparison

To our knowledge, there is no other approach that has mapped references through Wikipedia revisions, so there is no state of the art method that we can use to compare our method against. Therefore, we implemented a straightforward baseline that maps references using cosine similarity between numerical representations of the strings of the Gold Standard reference pairs (via Bag of Words representations) Then we resampled using the distribution of cosine similarities calculated in the original data. To estimate the distribution we used the same procedure of random sampling (Subsection 4.1) but we assume that the buckets have an infinite size (Appendix B, Step 1), and stop the algorithm after 100,000 pairs of references have been sampled. Figure 4 shows how our method leveraging WikiWho and Jaccard similarities outperforms the alternative based on Cosine similarity between reference strings through all possible thresholds.

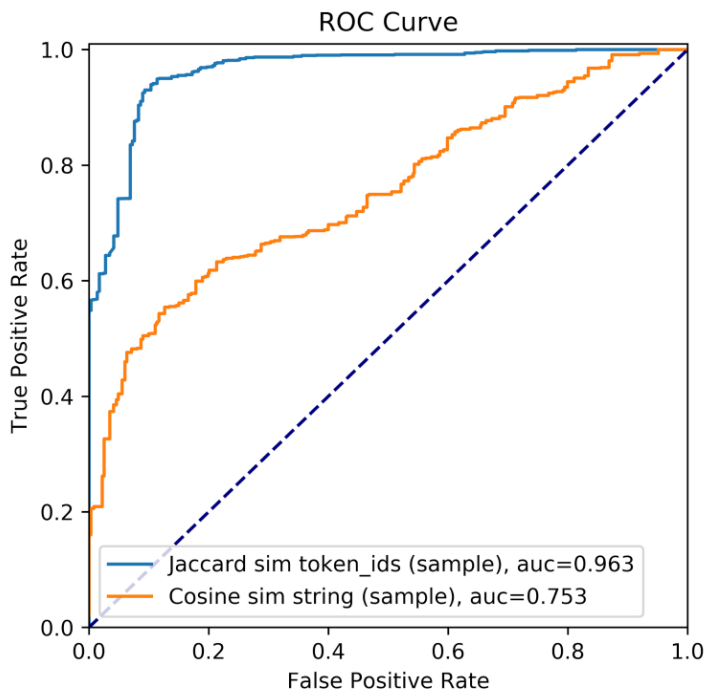

**Figure 4 - ROC Curves to compare our method and a simple method based on cosine similarity.** The light blue line shows the ROC curve for our method based on Jaccard similarity over WikiWho token IDs, and the orange line the ROC curve for a method based on cosine similarity of strings. Each data point is calculated for each possible threshold in the sample data.

## 5. Dataset composition and analysis

Our dataset contains the references of 6,073,708 non-redirect[24] articles in the English Wikipedia. It comprises 55,503,998 references with 164,530,374 actions. The actions are divided into 33.73% creations, 31.3% modifications, 23.15% deletions, and 11.81% reinsertions. We find that 77.21% of the articles (4,690,046) have at least one reference (median = 4, μ = 11.83, max = 12,797). But out of those articles, 78.42% do not yet have any DID-Rs (3.68 million, i.e. 60.54% of total articles; see Figure 5). The rest of the articles (1,012,289) have at least one DID-R, and 50,615 (5%) articles contain more than 50%  DID-Rs. More than 88% of the DIDs currently used to track the references correspond to ISBNs and DOIs (Figure 6).

---

[24] We excluded Wikipedia pages that are redirects. Redirects are Wikipedia pages that automatically send visitors to another page and do not have their own content. Example: https://en.wikipedia.org/w/index.php?title=Symbiont&redirect=no

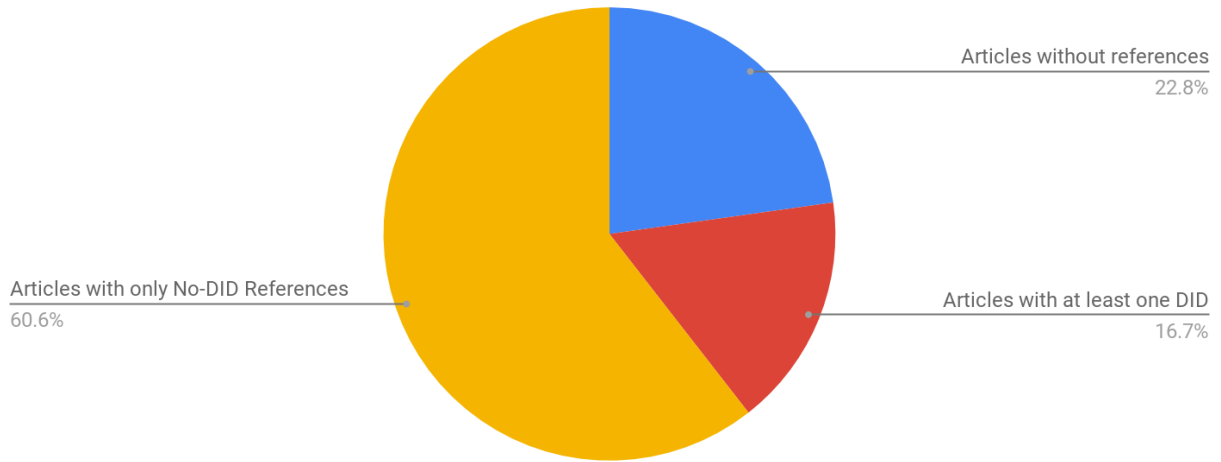

**Figure 5 - Distribution of articles according to the presence of references that include a document identifier (DID References, DID-Rs).**

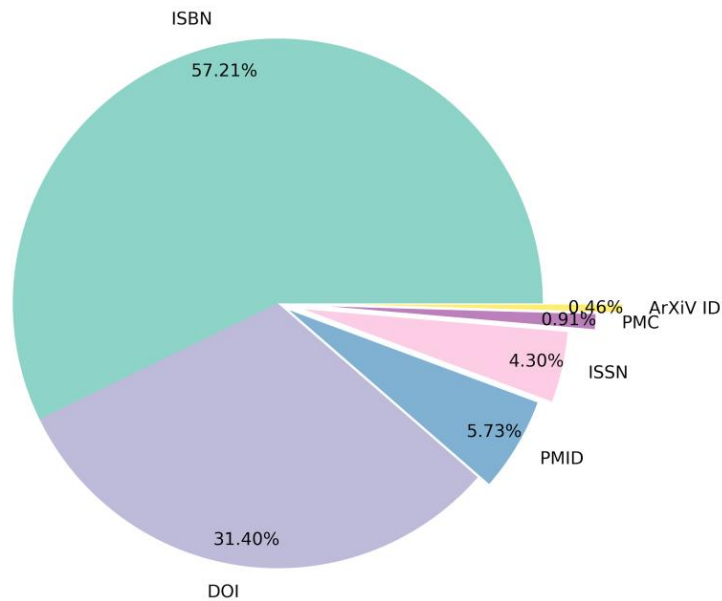

**Figure 6 - Distribution of DIDs by the type of identifier associated with the references.** The graph includes all the DIDs found in all versions of the references.

As of June 2019, only 7.00% (3,943,984) of all articles include one of the identifiers we were tracking. Figure 7 compares the distributions of all reference counts (left) with the DID-R counts (right) per Wikipedia article, both of them suggesting power law distributions. However, the curve for all references is more flattened than the one for the DID-Rs.

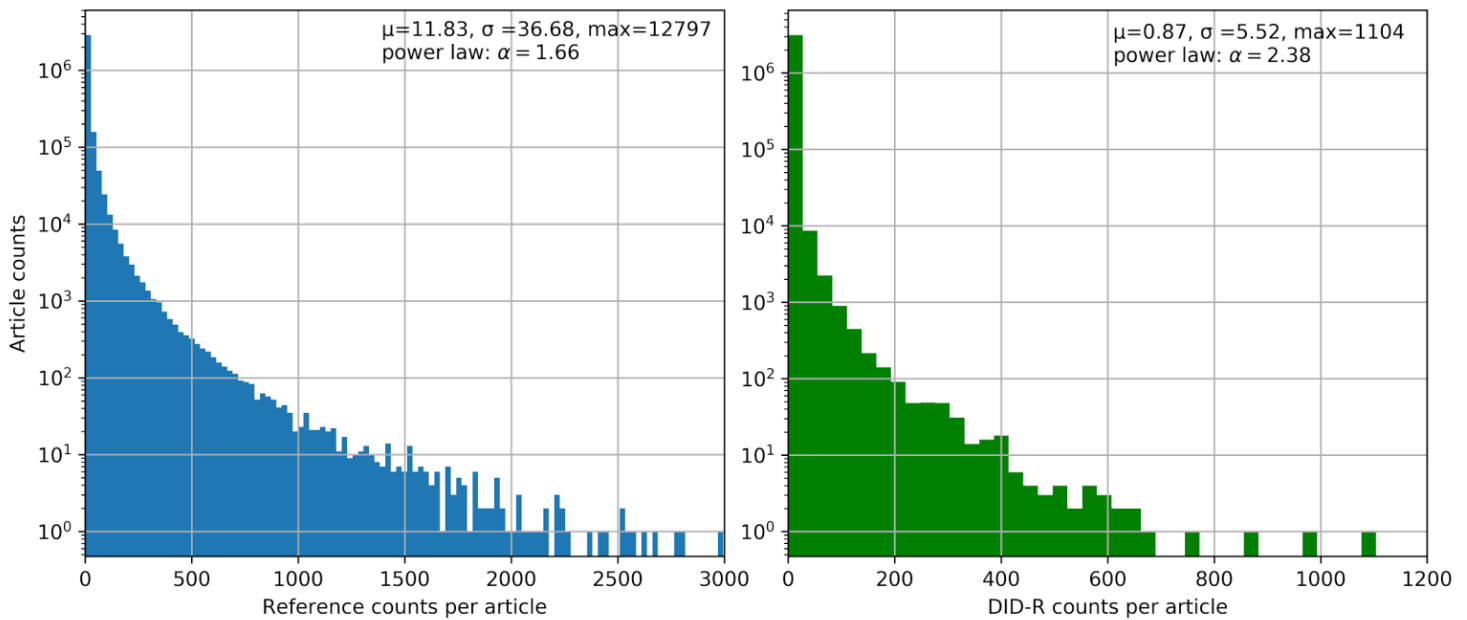

**Figure 7 - Distribution of reference counts in articles.** The X-axis aggregates the number of references per Wikipedia article into 500 bins (left) and 40 bins (rights). The Y-axis shows the number of articles using a logarithmic scale.

About 10% of all DID References are *DID-Lagged References*, i.e. they did not have DIDs in their early Wikipedia article revisions (Table 1). By now - and in the future - this number will likely be higher, as DIDs can still be added to the references that were classified as *No-DID References* in our 2019 dataset. We also observe that 12.1% of actions on the DID References occurred during the initial revisions in which the references did not yet have a DID; so this information would not be considered in any approach that relies only on DIDs for identifying and monitoring references.

In the following section, we will now take a closer look at the data in order to find some answers to our research questions. We will first look at the temporal evolution of different types of references (based on the presence of DIDs), and second on the editors who are creating and editing the references.

## 5.1 Wikipedia References over Time

The first reference in an article of the English Wikipedia edition was introduced in December 2005. Since then, more and more references have been added yearly (Figure 8). There was an initial steep increment of new references per year until 2010, in which more than 4 million references (which corresponds to 7.4% of all references) were created. After that, the increment of yearly created references continued more moderately, and it seems to have settled in 2017 and 2018: about 5.58 million (10.05%) and 5.64 million (10.15%) of all references were added in the respective years. To find out if this is a permanent trend, future investigations are needed.

After references have been created, some of them have never changed in any way, while others have been either deleted or modified at least once. According to our data, modifications are the most common action (~51.5 million) that happen to references after their creation. The number of modifications per year was not growing monotonically as we have seen for creations, e.g. there is a peak of modifications between 2016 and 2018: 6.41 million in 2016, 8.10 in 2017, and back to 6.71 in 2018. We suspect that the increase of modifications 2016-2018 is due to WikiCite[25] project and sequence of events that started in 2016.

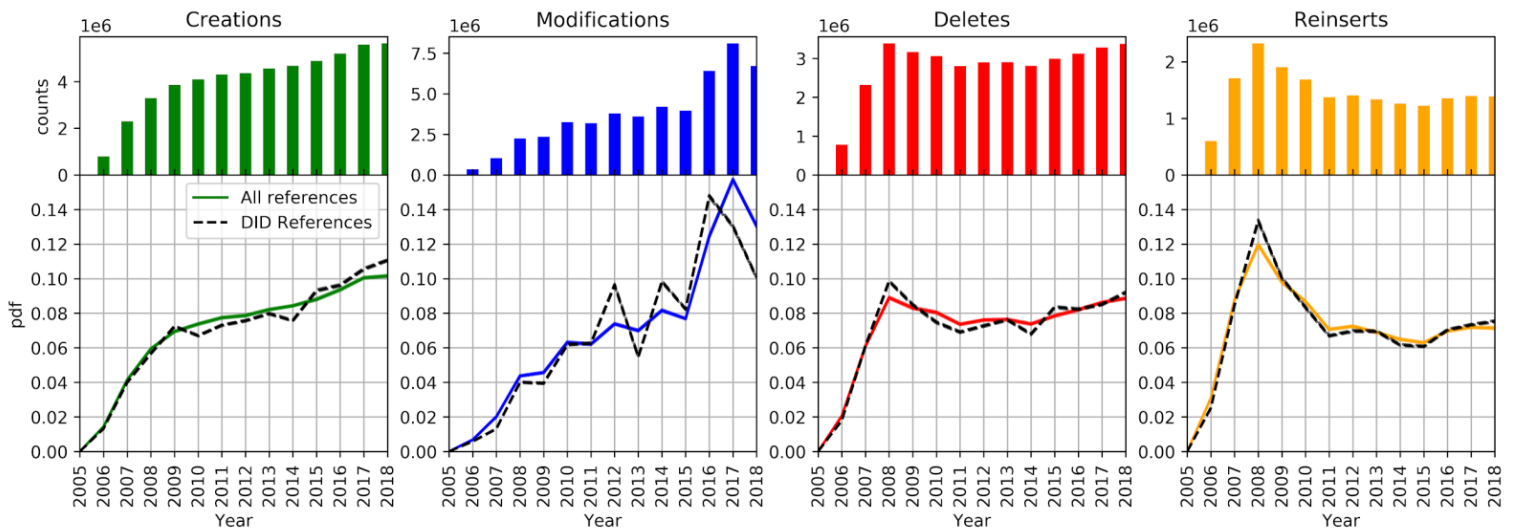

**Figure 8 - Distribution of actions over time.** Each of the four plots depicts the dynamics of one of the actions: creations, modifications, deletions, reinsertions. On the top subplot of each action, bars represent the number of actions (Y-axis) performed over all references per year (X-axis). For example, around 2.3 million references were created in 2007. On the bottom subplot of each action, the solid lines represent the proportion of actions (Y-axis) that occurred yearly (X-axis) for all references. The dashed lines represent the proportion of actions that occurred yearly (X-axis) for only the DID References (DID-Rs). For example, around 8.9% of all deletions have been done in 2008, whereas for DID-Rs around 9.9% of deletions have been done in 2008.

With the exception of 2005-2006 (years with small reference counts), the number of deletions has shown a decreasing trend until 2014. This was most likely due to clean-up efforts of initial reference additions, plus high volatility, e.g., because of disagreements, also shown in the high reinsertion

---

counts until 2010, which are by default a reaction to previous deletions.[26] Starting at its high count in 2008, the number of reinsertions was unevenly dropping from 2.33 (11.98%) million actions in 2008 till 1.39 (7.14%) million actions in 2018. Deletions have dropped since 2015, entailing a constant reference growth.

One might expect the same distribution of actions across years for DID References, i.e., that they would be treated by editors in the same way as general references. Yet, we can see some differences between general references (Figure 8, solid lines) and DID-Rs (dashed-lines). The most distinct patterns are noticeable in the creations and modifications of references (the dashed and solid lines, Figure 8):

- Until 2009 the amount of creations of DID-Rs was aligned with creations of all references (overlap of the dashed and continued line). However, between 2010 and 2014 fewer (than expected) DID-Rs were created, and after 2015 the trend was reversed. For instance, in 2018, around 11.06% of new DID-Rs (versus 10.15% of general references) have been added to Wikipedia articles.
- There is no clear trend in the modifications of DID-Rs (the second plot from the left, Figure 8), as the plot shows multiple peaks and troughs across the years. We observe fewer modifications of DID-Rs in 2007-2009, 2013, 2017, and 2018; and more modifications in 2012 and 2014-2016. The highest number of modifications have been reached in 2016 (1.02 million actions or 14.79%) and 2017 (0.9 million actions or 13.03%).
- The relatively small differences in deletions of some years (2008, 2010-2012, 2014, and 2015 in Figure 8) do not necessarily mean that their presence ended in those years (since they can be reinserted).

We found that DID-Rs have a higher survival rate: they are deleted (without further reinsertions) at a lower rate than the rest of the references. As of June 2019, around 31.8% (17.02 million) references have been deleted (without further reinsertions) between 2005 and 2019, 0.97 million of them are DID-Rs (i.e. 25.7% of all DID-Rs). Figure 9 presents the cumulative percentage of deleted references for each year, we observe that the percent of deleted references grew from ~20% in 2007 to ~32% in 2019 (and from ~11% to ~26% for DID-Rs correspondingly). This speaks to a higher value of these references to the editor community, possibly because of their perceived quality.

---

[26] The years 2006-2010 in the English Wikipedia have been pointed out as a highly volatile period before (Flöck et al. 2017)

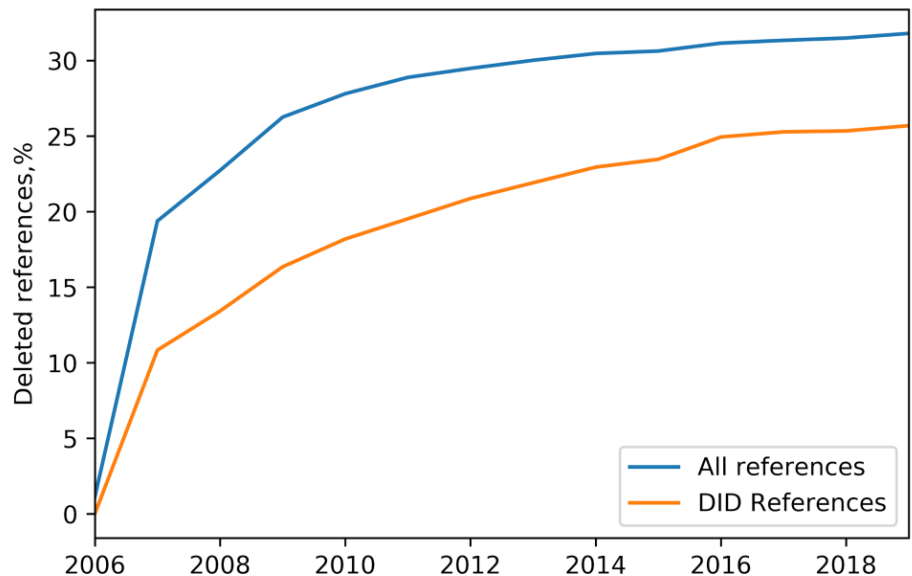

**Figure 9 - Percentage of deleted references**. The orange line shows the percentage of DID-Rs that were deleted (w/o further reinsertions) before any given year in the X-axis; the blue line shows the percentage for all references. For example, if we would analyse Wikipedia references as of 1 January 2010, we would observe that 27.8% of all references and 18.2% of DID References have been deleted without being reinserted before June 2019.

We observed in Figure 8 (second subplot from the left) that there are clear differences in the overall number of modifications, and the number of modifications of DID references. Some of these modifications are of particular interest because they are the ones in which DIDs are added to already existing references (by definition, references that were classified as DLag in Table 1). Therefore, we have closely investigated these modifications (Figure 10). We observed that the highest peaks of newly added DIDs occurred during: (1) May and June 2008 with 22,126 DIDs added during two months, (2) May 2014 with 18131 DIDs added, and (3) February 2019 with 12,486 DIDs added. Based on information until June 2019, these three peaks correspond, respectively, to (1) 19-26%, (2) 17%, and (3) 56% of references that at the time should have had a DID (see Appendix E for statistics of other peaks). Putting it the other way around, 44-83% of references remained without DID even after pronounced waves of DIDs additions.

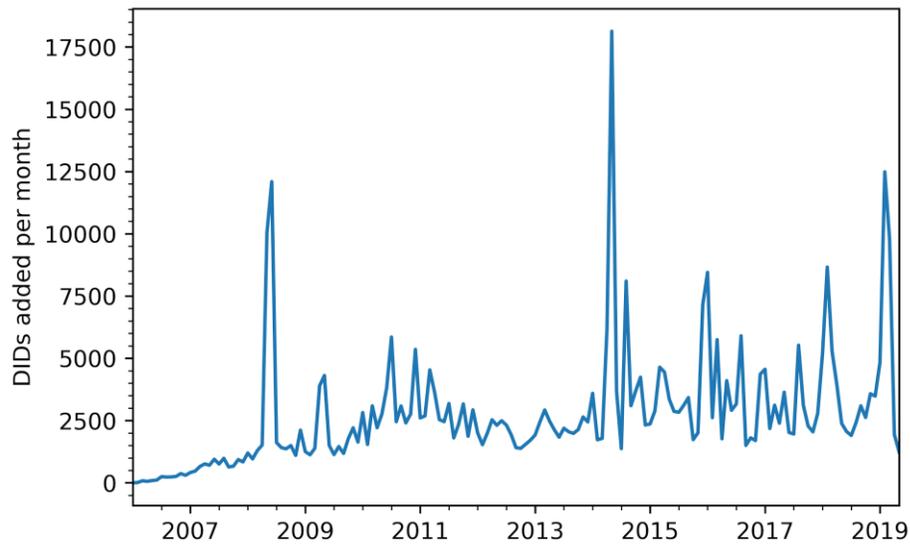

**Figure 10** - Monthly total of modifications that added a DID to existing references. The X-axis displays the year and y-axis the amount of modifications. Only modifications in which a DID was added to a reference are considered.

The latter percentages will be even higher in the future (after June 2019) as more DIDs will be added to references that existed at those peaks. Hence, we also analyzed how long it takes for the reference to be attributed with DIDs. Figure 11 presents the distribution of time span between reference creation and DID introduction for references created in three different years. In 2006, it took between 500 and 1000 days for most of the references. In contrast in 2018, it took less than 10 days for most of the references to get a DID.

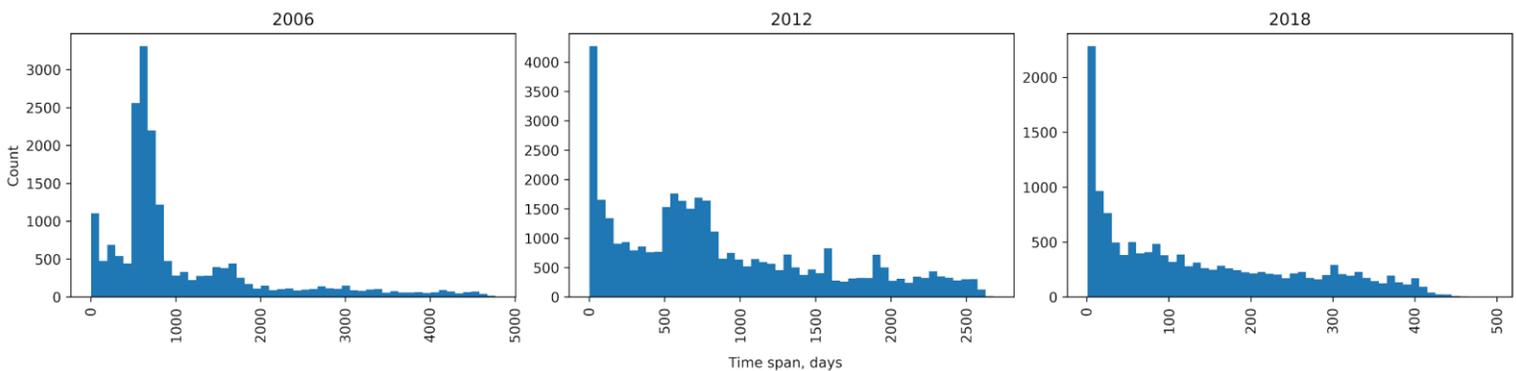

**Figure 11 - Distribution of the time spans between the creation of the references and the introduction of their DID for the years 2006, 2012, and 2018**. The X-axis shows the time span in days between reference creation and the introduction of their DIDs (only including the references created in each of the years in the titles of the plot). The Y-axis shows the frequency for each of the time spans. See Appendix D for distributions of all other years.

As we already mentioned, DID References correspond to 7% of all the references in our 2019 dataset - but the question is whether this would have been different at earlier points in time. Would we have collected the dataset in other years, the percentages would have been slightly different (solid line, Figure 12). For example, if one would ask the same question at the beginning of 2007, there would have been around 6.6% of DID References. After a recovery in 2010, the number of DID References has stabilized around 7% with a small increase in the last 4 years.

Hypothetically, one could collect the histories of references using only DIDs (see Appendix C). In that case, one would observe ~4.4% of DID References in 2007 (dashed line, Figure 12) where the true number should have been at least 6.6% references; the alternative method would have missed 37.5% (~2.2% out of ~6.6%) of references that got their corresponding DID after the hypothetical data collection. These differences are discussed in more detail in Section 6.

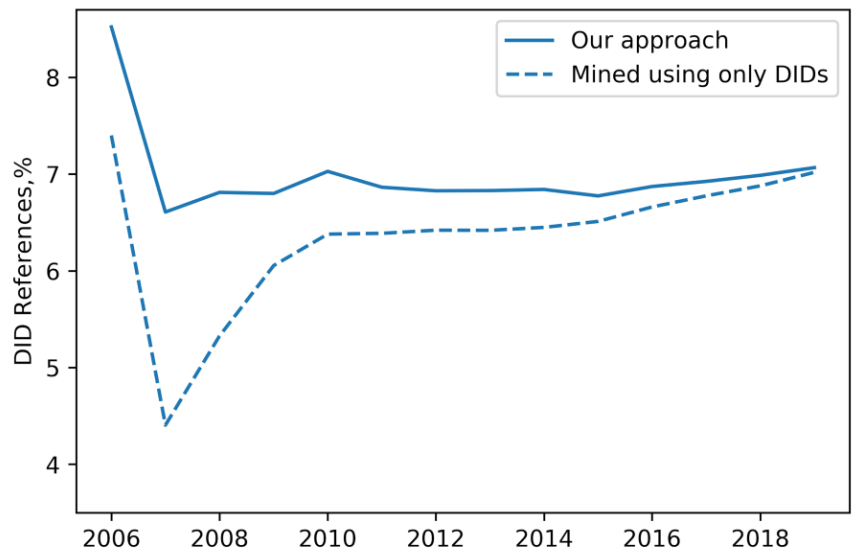

**Figure 12 - Percentage of DID-Rs at different time points**. The solid line defines the percentage of DID-Rs(Y-axis) at certain time points (X-axis) obtained using our approach, the dashed line defines the percentage of DID-Rs (Y-axis) at certain time points (X-axis) obtained using DIDs (with which we matched references between revisions; see Appendix C).

## 5.2 The Editors of Wikipedia References

In the context of altmetrics, the focus is often placed on which scholarly works receive mentions or interactions from social media or other alternative platforms, while relatively little is known about who is behind these mentions and interactions. In social media platforms such as Wikipedia, it is relevant to understand the actors who participate in the inclusion of scholarly publications as this has a direct impact on visibility. In contrast to traditional publications where the decision of which material should be cited is attributed to the authors of each publication, in a collaborative environment, the decision is not straightforward but may have to be negotiated over different article revisions. In this subsection, we investigate (1) whether contributions come from registered editors, bots, or non-registered sessions (IP addresses) (see Table 3), and (2) explore the behaviour of these actors within Wikipedia. In particular, we are interested in whether those who edit Wikipedia references differ from the overall Wikipedia editor community, and we inquire if there exist g sub-communities of editors that specialize in different types of editing activities (creation, deletion, modification, reinsertion).

**Table 3 - Types of Wikipedia Editors.** The first column lists the types of editors, the number of reference-editing actors of each type and their actions that we encounter in our dataset. The second column elaborates on each.

| Types of Editors | Description |
|---|---|
| **Registered Editors**<br>Actors:        1,910,667<br>Actions:    121,681,174 | These correspond to individual users who have registered their profile on Wikipedia and edited at least one reference. |
| **Bots**<br>Actors:              1,172<br>Actions:      19,386,851 | Bots were identified from bot lists of Wikimedia plus an additional list of bots' names that we created. These sources were combined into a final list consisting of 10,262 unique account names (see Appendix F for the sources). Since Wikipedia has strict rules and mechanisms to combat spam and any automatic-like activities, all bots have to be registered accounts. |
| **Non-registered Editors**<br>Actors:              N/A<br>Actions:      23,459,838 | Edits coming from non-registered IP addresses cannot be attributed to specific anonymous editors. Several persons can share the same IP address (e.g., university addresses or libraries), and one editor can connect via several IPs. |

We found 1,910,667[27] editors, 1,172 bots, and 23,459,838 edits by 4,286,160 IP addresses that worked with Wikipedia references (Table 3). Figure 13 presents the distributions of actions per user type. Registered editors are responsible for the majority of actions, i.e., more than 122 million (74% of all actions in our dataset). Registered editors focused on the creation of new references (40% of their actions) and modification of existing ones (28.2% of their actions). Bots, in comparison, with a total of 19.4 million (13.7%) of all actions, were focused on modifications (that corresponds to 71% of their actions). Non-registered editors are responsible for only 14.3% of the actions in our dataset. And although registered editors made most of the deletions (around 24.5 million, left plot in [Figure 13](#)), non-registered editors appear to specialize on them (right plot): non-registered sessions have proportionally more deletions (53.3% of all their actions) than either registered editors (20.2%) or bots (5.6%), not unlikely due to large amounts of vandalism, especially blanket deletions of large chunks of text. The non-registered editors seem to comprise a very diverse and occasional set of editors, as 89.8% of IPs have less than 10 actions at all. This figure could be higher when we consider that some IPs might be associated with several editors (e.g., school IPs); we assume that it is unlikely that one individual would use a large number of IPs. Given the comparably low figures of actions for non-registered editors but mostly the difficulties of attributing actions to specific actors, we will exclude them from the rest of the analysis in this section.

---

[27] For comparison, English Wikipedia has a total of 35.7 million registered editors as of July 2019.

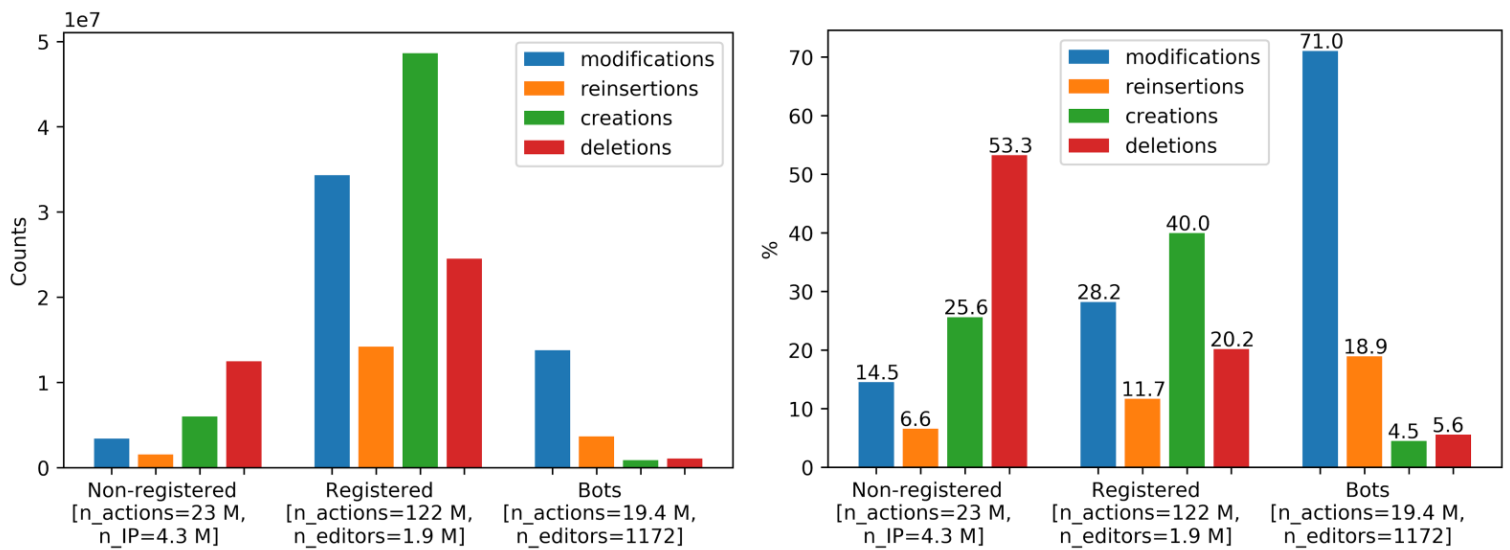

**Figure 13 - Distribution of actions performed by type of editors.** The left plot shows the total actions (Y-Axis) per type of account (X-Axis), and type of action (legend). The right plot shows the percentage (Y-axis) of the type of actions (legend) within the account type (X-axis). The X-axis also presents the total number of actions (n_actions) and editors or IP addresses (n_editors or n_IP) for each account type.

Most registered editors have performed only a few actions on references in Wikipedia articles, whereas the top-contributors have contributed millions of action ([Figure 14](#)). We also studied the number of different articles in which each editor has performed actions on references. We see a similar trend as with the number of actions, the user at the top has edited references in 226,334 articles. This seems to suggest that some editors are specifically focusing on reference editing beyond a specific topical area of interest.

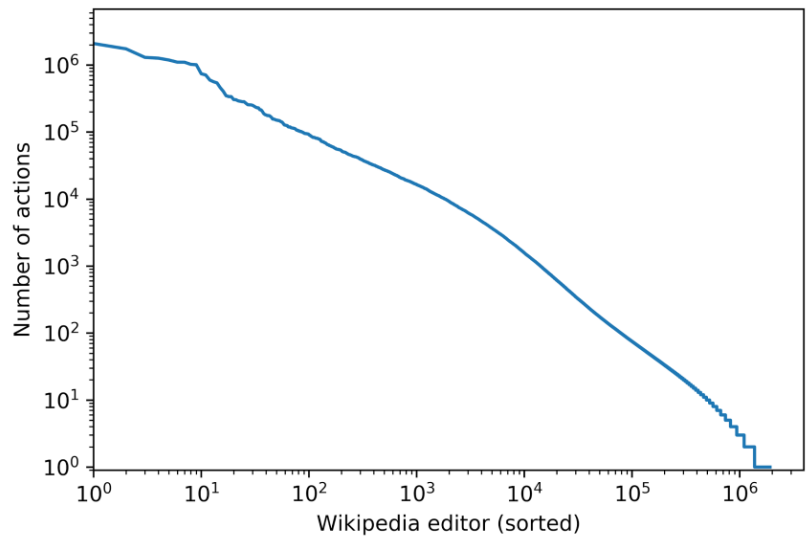

**Figure 14 - Distribution of actions on Wikipedia references per registered editor.**

Using a manually curated list of Wikipedia bots (10,262 unique bot account names, see Table 33), we found that 1,172 bots (0.1% of editors) have taken part in the edition of references. On a per user basis, bots performed more actions on references than registered users (Mann-Whitney U test, p<0.001; Figure 15).

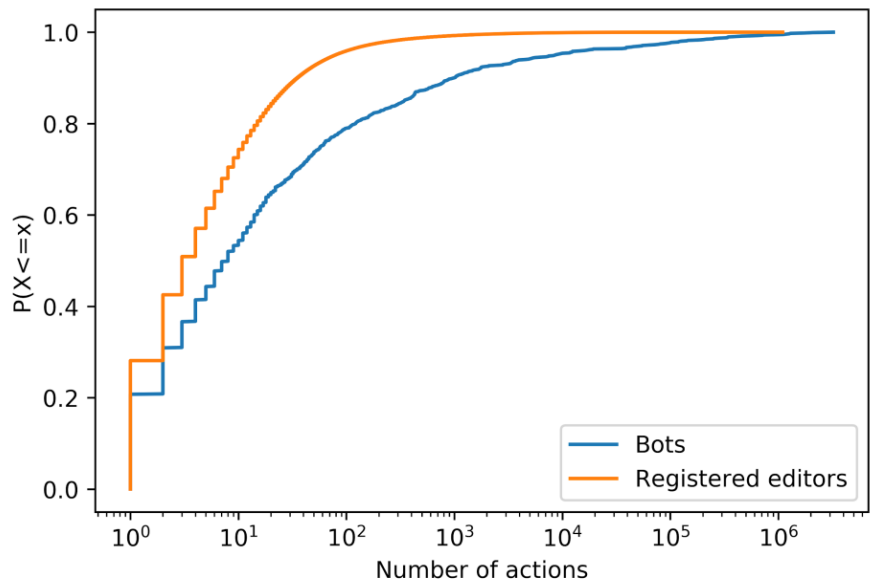

**Figure 15 - CDF of actions performed by bots and registered editors.** The X-axis shows the number of actions (x), and the Y-axis the probability of a user having less than x actions. For example, 85% of bots have less than 1000 actions, whereas 85% of registered editors have less than 30 actions.

Bots and registered editors display very different behaviour that is evident by directly looking at the types and quantity of actions (Figure 13 and Figure 15). Within the group of registered editors, we were interested in identifying subgroups of users who behave similarly (and distinct from other sub-groups), as measured by the types of actions that they usually perform.

We use the K-means clustering algorithm to find such groups. Each registered editor is represented by 4 features, one per type of action, that contain the distribution (in percentages) of actions of that editor. We applied the algorithm on a sample of 10,000 random editors. To determine the optimal number of clusters the following analyses were performed: (1) silhouette coefficients (Rousseeuw, 1987), and (2) Clustering tree algorithm (Zappia & Oshlack, 2018) with Sugiyama layout (Sugiyama et al., 1981) for tree depiction.

The silhouette measures how close each editor in one cluster is to editors in the neighbouring clusters. For example, classifying editors using 6 clusters (k=6) resulted in a mean silhouette score 0.69 (Figure 16): (1) editors from clusters 0, 1 and 2 (with mean silhouette score between 0.9 and 0.95) are well separated from other clusters as all the editors have a positive silhouette and most of them a very high one (>0.85), and (2) clusters 3, 4 and 5 have outliers because some of the editors have a negative silhouette (indicating that they are very close to other clusters.

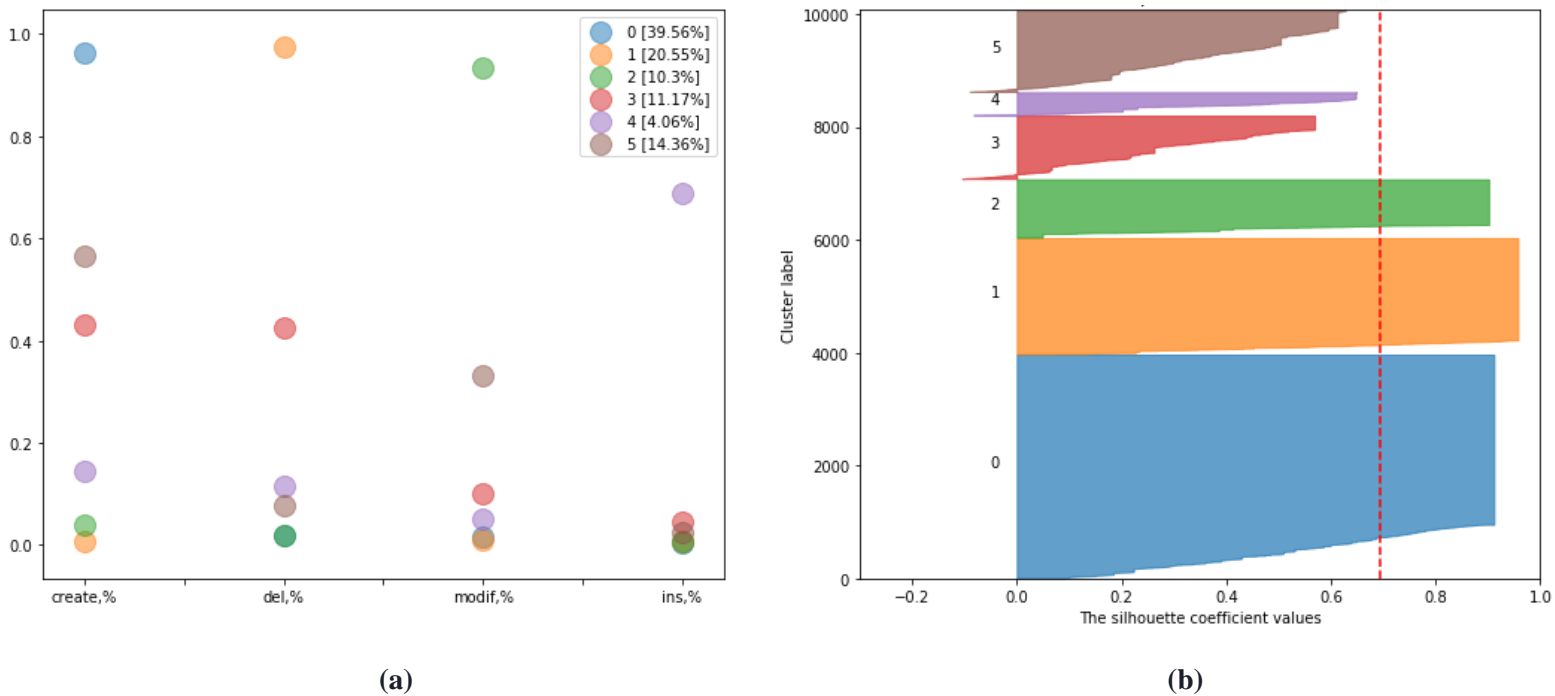

(a) (b)

**Figure 16 - K-means clustering results (k=6) for clustering registered editors by types of actions.** a) centroids of clusters, b) the silhouette coefficients of clusters

The editors were classified using different values of k (from 1 to 11), and then a tree was created ([Figure 17](#)). By traversing the tree from top to bottom, we can observe the composition of each cluster at a given k with respect to the cluster of the previous level. For example, cluster 6 at k=7 contains elements of clusters 0, 2, and 5 from k=6. Therefore it is likely a non-well-defined cluster (unstable). This suggests that stopping the division of clusters at k=6 would provide us with a more stable configuration of clusters. This observation is confirmed by the average silhouette scores.

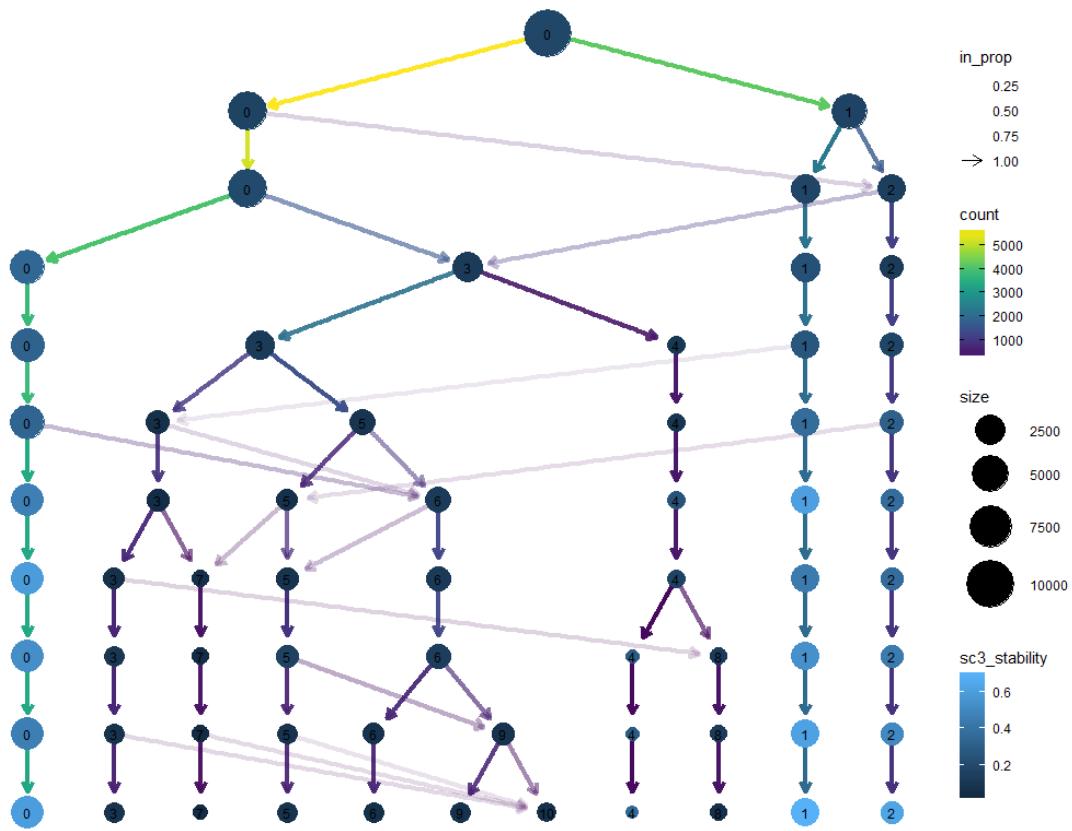

**Figure 17 - ClusTree of reference editors.** Each level of the tree (from top to bottom) corresponds to the k used, and each node on that level corresponds to a cluster of that k. The edges (arrows) of the tree represent the editors that "move" from one cluster in level k to another cluster in level k+1. The legend of the graph displays 4 scales, from top to bottom: (1) the transparency level of arrows (in_prop) shows the proportion editors from one group that end up in another group. (2) The arrow colour (count) shows the number of editors that "move" from one cluster to another, (3) the node size is proportional to the number of editors in the clusters, and (4) the nodes colour intensity depicts the stability index (Kiselev et al. 2017).

With the 6 clusters, we clearly observe different behavioural patterns that characterize the registered editors. Table 4 summarizes this characterization.

**Table 4. Classification of registered editors according to the type of activity.** The first column presents the cluster id and the percentage of registered editors in parenthesis. The second column describes the group of registered editors in terms of the actions they perform (Figure 16a).

| Cluster (%) | Type of activity |
|---|---|
| 0 (39.56%) | Only create new references |
| 1 (20.55%) | Only delete references |
| 2 (10.3%) | Modify references in 90% of the cases, create new ones in 6% cases |
| 3 (11.17%) | Mostly delete and create references (42% of cases for each action), modify in 10% cases and do a few reinsertions |
| 4 (4.06%) | Mostly (70%) reinsert deleted references and do a few deletions, creations and modifications |
| 5 (14.36%) | Mostly create (55%) and modify (35%), and do a few deletions and reinsertions |

Additionally, we wanted to know whether editors of references are different from the general Wikipedia editor community (Wikipedians). We have therefore compared the 10,000 most productive reference editors in our dataset with the most productive Wikipedians according to Wikimedia Foundation[28]. The ranking of the most productive Wikipedians is based on the total number of revisions they have created, whereas we have used five different rankings of reference editors based on the following criteria: (1) total count of actions, (2) modifications, (3) creations, (4) deletions and (5) reinsertions. To see if highly productive reference editors correspond to highly productive Wikipedians, we utilized rank-biased overlap (RBO) by Webber et al. (2010). Apart from accounting for the position in the rank of each editor, RBO properly deals with two characteristics of our rank comparison task: indefiniteness (the rank range of 10,000 is arbitrary) and top-weightedness (the variation among the top-active editors is more relevant than the one among the rest) of the editors. Regarding the latter, the weight is determined by the research using the parameter p (ranged from 0 to 1); the lower it is, the more importance is placed on the top results.

---

[28] https://en.wikipedia.org/wiki/Wikipedia:List_of_Wikipedians_by_number_of_edits

We calculated different values of p ([Figure 18](#)) as a way to explore, and found that the RBO scores are very low for values below 0.95 which place a lot of importance to the order; we only start seeing higher scores after values above 0.995 (RBO Scores ~0.016). The low scores cannot be attributed to the lack of common editors between the rank lists (of 10 000 editors), as scores go relatively high with extremely high values of p past 0.99995.

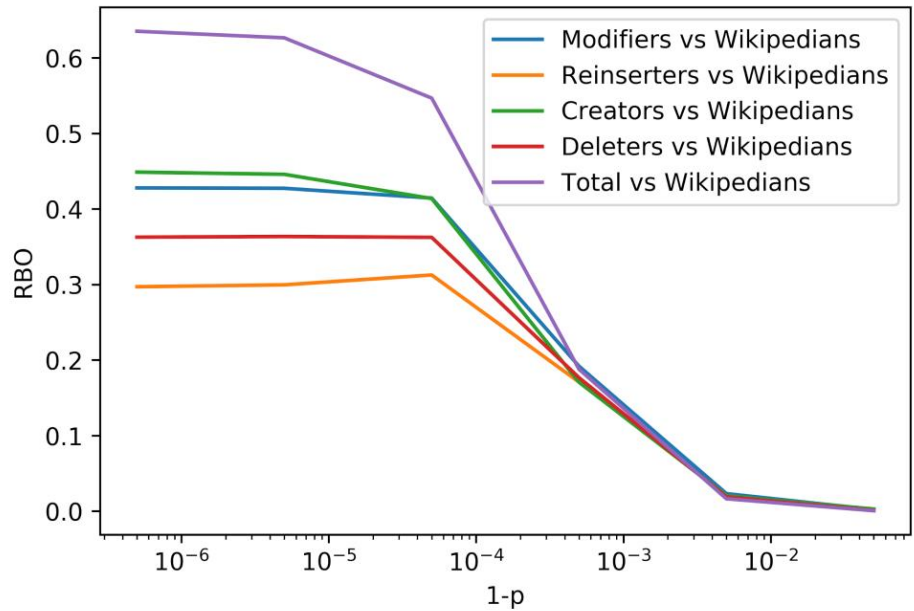

**Figure 18 - RBO scores between ranked lists of productive Wikipedians and the most active editors of a certain action.**

To clarify this effect, we also present a more intuitive index, Jaccard similarity ([Table 5](#)), which simply takes the magnitude of the intersection of elements (editors) between two lists divided by the total number of different editors. For example, 52% of editors are in both, the top-active Wikipedians and the top-active registered editors of all actions in Wikipedia references. Not only is the latter group very different from the normal Wikipedians, but also the lack of overlap is more notable for groups specialized in certain types of actions. For example, the cluster 1 ([Table 4](#)) dedicated to only deletions has a Jaccard index of 0.42 (for the top 10,000) and an even lower RBO score (0.363 with p of 0.9999995); even more distinctive is the group of editors doing reinsertions (last row of [Table 5](#)), which well describes cluster 3, i.e. editors with mostly reinsertions.

**Table 5** - **Similarity of action groups with productive Wikipedians.** The first column (actions) displays the criteria used for our ranking. The second column represents the RBO scores for two values of the parameter p of RBO. The rest of the columns show the Jaccard similarity of first top-x editors of both rankings.

| Actions | RBO scores for p=[0.95, 0.9999995] | Jaccard similarity of top | | | | | |
|---|---|---|---|---|---|---|---|
| | | **10** | **100** | **500** | **1000** | **5000** | **10000** |
| Total | [0.001, 0.635] | 0.05 | 0.25 | 0.32 | 0.38 | 0.48 | 0.52 |
| Modifications | [0.002, 0.428] | 0.11 | 0.18 | 0.25 | 0.29 | 0.42 | 0.46 |
| Creations | [0.003, 0.449] | 0.00 | 0.12 | 0.20 | 0.26 | 0.37 | 0.41 |
| Deletions | [0.001, 0.363] | 0.11 | 0.17 | 0.25 | 0.31 | 0.41 | 0.42 |
| Reinsertions | [0.001, 0.297] | 0.05 | 0.09 | 0.16 | 0.20 | 0.34 | 0.37 |

# 6. Discussion

With this work we have presented a high-quality dataset of Wikipedia references (prepared by utilizing a new method also introduced and evaluated here) and have used this dataset for initial investigations into the edit histories of Wikipedia references and the role of different types of Wikipedia editors. The dataset can be used and repurposed by the scientific community in the future.

## 6.1 Quality and applications of the dataset

To the best of our knowledge, we have created the most comprehensive dataset of English Wikipedia references to date. To achieve this, we mine all these revisions, and did not limit the process to only references that contain document identifiers (DIDs). Moreover, the dataset preserves the traceability of each reference across the revisions, including its creations, modifications to its content, and full deletions and reinsertions of the text corresponding to the reference. Our evaluation indicates that our method has an accuracy (based on F1 score) above 96%[29] against a gold standard based on judgements by crowdworkers that we also contribute as part of this work. At the same time, since we manually inspected the false positive cases of our system, we are certain that the quality of the gold standard is very high and could be used to evaluate

---

[29] The reported F1 score corresponds to the method that was calculated using stratified samples, but since the distribution of similarity of the references is biased towards 0 (completely different) and 1 (completely similar) the accuracy of the our dataset is higher

future methods, as we did with the baseline method based on cosine similarity (Subsection 4.3). Researchers could compare our method to other bibliometric matching algorithms, such as ones provided by the Centre for Science and Technology Studies (CWTS), the Institute for Research Information and Quality Assurance (iFQ), or Web of Science (WoS)[30] (Olensky et al., 2016). However, we argue that those methods would not perform well as they depend on bibliographical fields, which are often missing in Wikipedia references, not to mention the existence of abundant errors, especially in early revisions.

Despite our efforts to ensure the high quality of our dataset, we had to make certain compromises that lead to the following limitations: (1) While our method covers references indicated as inline citations via ref tags across their edit histories, we do not include any other forms of references that coexist in Wikipedia, e.g., parenthetical references[31] or wikilinks to full references using templates[32] outside of ref tags. This decision was partly because the format of these references is not uniform and we cannot guarantee that their extraction would be accurate, but it is also in line with Wikipedia's recommendation for how references should be added to articles. The latter also implies that we can assume some quality control for inline citations based on Wikipedia's standards. Altmetric.com is now also only considering ref tags for accessing references and argue that other forms like sources mentioned as additional reading can be easily manipulated. (2) The dataset was created based on the English Wikipedia as of June 2019. We do not know anything about edits that happened after this point in time. In general, this causes that the estimates of DID coverage are optimistic because we do not know if more DIDs were (or will be) added after that date. (3) We worked with a selection of common types of document identifiers that we summarized as DIDs. For our analyses that include information about whether a reference included a DID at some point in time, this only pertains to the following types of identifiers: DOI, PubMedID, PMC, ISBN, ISSN, ArXiv ID. The list of identifiers is the same used by the Wikimedia Foundation project (Halfaker et al., 2019), as it might capture most academic citations. We use the presence of identifiers as a weak indicator for quality of the referenced publications, and not as a clear characteristic to distinguish between different types of publications, e.g. scientific vs. non-scientific. (4) For the clustering in Subsection 5.2, we did not take into consideration anonymous editions to Wikipedia because we only have access to the IPs of the editors, and it would be wrong to assume a one-to-one relation between an IP and an editor.

---

[30] http://apps.webofknowledge.com

[31] https://en.wikipedia.org/wiki/Wikipedia:Parenthetical_referencing

[32] For example, shortened footnote template (e.g., {{sfn}}), Harvard style templates (e.g., {{harvnb}}), or freehand anchors (e.g., [[ #anchor_id]])
https://en.wikipedia.org/wiki/Wikipedia:Citing_sources/Further_considerations#Wikilinks_to_full_references

In spite of the limitations, we hope that by providing our dataset to the research community, we contribute to future work in the area of altmetrics. For example, this can be used to compare different data collection approaches (e.g. used by altmetrics aggregators), or to evaluate previous datasets used in altmetrics publications against one that include the historical evolution of those existent references. It also opens the opportunity to investigate additional research questions related to coverage of specific types of publications over time, background information for evolutions of highly-cited publications, topical distributions of references, and the dynamics of editors that surround the references. We already start this work by providing insights into the evolution of references based on edit types (actions), DID coverage, and editor characterization.

## 6.2 Evolution of Wikipedia references

For our first research question (RQ1) we were investigating in more detail how Wikipedia references evolve over time. Our data clearly highlights that references in Wikipedia have to be viewed as a continuum that evolves in different dimensions. These insights should not be underestimated when working with Wikipedia data in the area of altmetrics, as they imply that the point of data collection will be crucial for any observations and that ways are needed to also account for vanishing or changing references: citation counts for publications based on Wikipedia data will thus not only increase, but may as well decrease over time. Between 19.4% and 31.8% of total references (between 10.8% and 25.7% of DID-Rs) were deleted every year (from 2007 to 2019) and never reinserted again. These full deletions could cause erroneous assumptions drawn from statistics or even correlation analyses, and imply an instability of Wikipedia mentions as a measurement instrument. In classical bibliometric instruments, comparable issues are negligible since changes of the reference list in papers are almost impossible, and retractions of paper (together with their referencing lists) are very rare events (Shema et al., 2019). But for altmetrics, the phenomena of citation data volatility needs to be discussed in the community.

Besides the quantification of different types of actions as presented in Section 5, we were able to observe additional tendencies of reference evolution. Regarding RQ1, we find evidence that there is a continuous effort to increase the quality of Wikipedia references, expressed in the following ways: First of all, the number of references added to Wikipedia is constantly rising. Another evidence, regarding the effort to increase the quality of references, is the sharp increase of modifications in the last three years.

Interestingly, we found that in some of the years there are peaks of modifications only targeting DID-Rs. We believe that these peaks represent particular efforts to add missing DIDs as the peaks are often followed by low values (troughs) - probably because there is a decrease in the amount of missing DID-R that can be detected by the normal editor community. This salient pattern shows that DID-Rs are treated very differently by the Wikipedia community.

Assuming that the presence of DIDs is an indicator of credibility, the proportion of new references that already include a DID (DBorn) has been consistently higher in the later years. The Wikipedia editor community seems to also perceive the reference with DID (DID-R) as more credible, as we conclude from the fact that DID-Rs are deleted with lower rates than all references.

## 6.3 The role of identifiers and potential effects for altmetrics

Not only because of the discovered differences in the ways Wikipedia editors treat DID-Rs, but also because of the general importance of document identifiers (e.g. for identifying and tracking publications that were cited by Wikipedia articles), we placed an additional focus on the evolution of document identifiers as elements within Wikipedia references (RQ2). We argue that changes performed on DID-Rs have a higher relevance for those altmetrics tools that rely on identifiers to trace citations.

Full deletions of references might be the most intuitive case of disruption for measuring impact based on Wikipedia references, and it affects references with or without references in the same way. Modifications of references do not always have a direct implication for the altmetrics field: The only modifications that are relevant are those that change the reference in such a way that (1) they make it point to a new resource (i.e. equivalent to remove and add a reference), or (2) they make the reference either detectable (by adding a DID) or invisible (by removing the DID). The practical implications of these two relevant scenarios for the generation of altmetric data depend on the altmetrics mining method. But we assume that currently some of the altmetrics aggregators take advantage of the presence of DIDs for identifying and counting references from Wikipedia. Therefore, we have looked at the specific modifications that introduced a DID to an existing reference in more detail.

Specifically, we analyzed the DID-Lagged References that did not include a reference upon their first introduction, but received it through later edits. Those references would potentially have been ignored during their initial lifespans before getting their DID. We were able to show that they correspond to a considerable fraction of DID-Rs (10% corresponding to 12.1% actions before the introduction of the DID). We found important periods regarding the evolution of the DID-Lagged References ([Figure 12](#)). Before 2010, a method that would have only relied on DIDs would have missed up to 37.5% (2007) of references for which we know that they should have had a DID (as we see that their DIDs were added by June 2019). The situation quickly improved between 2009 and 2010 (11.3%), and then continued doing so until our data collection. As mentioned in the limitations, our estimates are optimistic as we cannot consider references where identifiers will be added after June 2019. Our findings show that mining methods that rely on DIDs are vulnerable to coverage errors (Sen et al., 2019), that can misrepresent the importance of academic works in the altmetrics community.

## 6.4 Towards understanding who edits Wikipedia references

In our last research question, we investigated the editor community that creates and maintains Wikipedia references (RQ3). Our findings can help to better understand who contributes to the body of references in Wikipedia. These contributors play a crucial role within Wikipedia as they judge the relevance of references and shape what Wikipedia readers may consume, also influencing whether an article is perceived as relevant based on the presence of references. This influence goes beyond Wikipedia, e.g. in the area of altmetrics, where citations from Wikipedia are contributing to different types of indicators that aim to measure publications' impact.

We found that most of the references (87.6%) are created by registered editors, whereas bots were only responsible for 1.6% of new references. The concern that the presence of bots that add references (Nielsen, 2008) was dominating the reference creation is not currently a generalized issue. For comparison, this has been the case in Twitter, where Robinson-Garcia et al. (2017) found that bots (and thoughtless bot-like retweets of user accounts) were responsible for most of the activity containing scholarly articles. These findings support the idea that Wikipedia references are curated by humans and thus involve deliberate selection of sources and materials; an essential feature to justify their use in altmetrics research.

While the learning that Wikipedia references seem to be largely maintained by registered editors, this does not yet tell us whether they are representative of the core Wikipedia editor community or whether some editors seem to explicitly specialize on reference editing. According to our similarity metrics (RBO and Jaccard), registered editors that participate in the evolution of Wikipedia references are considerably different from the rest of the Wikipedia editors. Furthermore, we were able to identify clusters of editors with very clear boundaries; two of these clusters are fully specialized in creations and deletions and together add up to ~61% of the editors. We also found single editors that edited references in many different Wikipedia articles (e.g. one editor has edited references in more than 226,000), and thus appeared to be highly specialized on reference editing independent of topical domains.

These observations deserve additional attention in the future, as they should remind us of our illustrative, introductory example ([Figure 1](#)). Despite Wikipedia being a community effort, individual persons can have substantial influence over certain areas. A single editor has the potential to largely affect the representation of a specific reference by adding (or removing) it multiple times. Non-registered editors could also be responsible for these types of exploitation of Wikipedia, at least we found that many deletions have been done by non-registered editors.

Alternative explanations for this observation could be that (1) registered editors might log out from their accounts to perform deletions of references (as a way of keeping themselves anonymous and free from possible repercussions of known peers), or (2) it might represent possible larger scale vandalism (i.e. attempts to delete big sections of Wikipedia articles) that escaped the filters of vandalic revisions included by the WikiWho service that we used. Further exploration is necessary to disclose the magnitude of the potential issue, but we highlight the value of our dataset to analyse this issue.

## 7. Conclusions

In this paper, we have introduced an overview of the evolution of Wikipedia references, analyzed the historical coverage of reference-mining methods that are based on DIDs, and offered a characterization of the Wikipedia editors. In the scope of our research questions, we conclude that the quality of Wikipedia references has been slowly but persistently increasing. Although our findings do highlight limitations, we believe that the historical registry of Wikipedia contains information that can be leveraged to create more robust methods of mining and assigning importance to references in Wikipedia. We recommend that such methods use this record to reduce manipulations and biases that blur the visibility of references, to increase the overall coverage of references (by looking at all revision), and to assign impact based on historical activity and the community of (e.g. reputable) editors that surround the references.

These recommendations only lighten up a different path for the creation of altmetrics based on Wikipedia, and there is certainly more to be done. The high-quality dataset that accompanies this paper offers the opportunity to extend the research in this direction, for example by (1) analyzing the longevity and activity of references distinguishing between academic and non-academic (see Singh et al (2020) for a classification approach), (2) exploring the dynamics of references according to different knowledge fields, (3) further investigating the editors by mining (with natural language processing techniques) their profile pages and extract demographics, (4) modelling the co-editors network to find important actors and communities, and (5) predicting which references are still missing a document identifiers since our dataset already provide this information for existing references.

## Acknowledgements

This research was supported by the Deutsche Forschungsgemeinschaft, DFG, project number 314727790. We would like to thank all the *metrics project members and also: Prof. Dr. Isabella Peters and Prof. Dr. Claudia Wagner for their supervision and feedback, student assistants Tara Morovatdar and Alexandra Stankevich for their help with the data curation and Kenan Erdogan for the insights about WikiWho service.

# Supplementary materials

**Appendix A - Crowdworkers' Task on FigureEight**

To validate our method, we created a gold standard dataset labelled by annotators. See [Section 4](#) for more details. Figures A1-A4 depict instructions that were presented to crowdworkers on FigureEight platform. Figure A5 shows an example question from the task.

With this task, we pay crowdworkers at least German minimum wage (9.19 Euro per hour as of April 2019). We estimated that for one Page with 6 Questions one crowdworker would need approximately 2 minutes. Thus, one worker can finish approximately 30 Pages in an hour and earn 10.50 US$ with the payment of 35¢ per Page.

## Overview

Your task is to **indicate if two Wikipedia references refer to the same resource**. A resource could be a book, news article, journal publication, website or any other source of information (online or offline). "Same resource" means that independently of representation or formatting, a reference refers to the same book/news article/journal publication/etc.

For example, the same journal publication can be represented in any of the following ways:

- Using **plain text**:
  *"Colby ,D . , et al . ( 2011 ) Prions .doi: 10 . 1101 / cshperspect . a006833"*
- Using a **URL**:
  *"http : / / doi . org / 10 . 1101 / cshperspect . a006833"*
- Using **Wikipedia markup** language (where attributes of references, like "title", are coded):
  *"journal | title = Prions | author = D. Colby | year = 2011 | doi = 10 . 1101 / cshperspect . a006833"*

*All the above-mentioned examples refer to the same journal publication ( i.e., "resource") with the authors being "Colby, D. et. al" , the year of publication "2011", the DOI (a unique identifier for publications) being "10 . 1101 / cshperspect . a006833" and the URL "http : / / doi . org / 10 . 1101 / cshperspect . a006833"*

## Rules

1. Identifying numbers like **doi, pmid, pmc, isbn, arxiv, bibcode, biorxiv, citeseerx, jstor** are the most reliable identifiers of the same resource.
2. The combination of title and author names are the second-most reliable identifiers of the same resource.
3. Web-Addresses (URLs) are good identifiers of the same resource if the identifiers under *Rule 1* + *Rule 2* are not provided. Certain limitations apply (see "*Hints*").
4. URLs are good identifiers of the same entities if titles and/or author names are not provided. However, not all references have URLs, or it can be missed in one of them.
5. References can be of different length, order and structure and still could reference the same resource.
6. References can omit many details (e.g., year), and still could reference the same resource.
7. References that refer to the same publication (e.g. a book) but *different pages in it* are to be considered the same resource.

## Hints

- URLs would usually help to identify the same resources. However, not all references have URLs, and sometimes references are represented by several URLs, including web-archive URL. Simply select the"Same resource" if at least one URL is shared.
- You can follow the link "*check the revision on Wikipedia*" below each reference pair, in order to find original URLs and how the references were modified and represented in the article.
- The differences between the two references (i.e., added and missing words) are highlighted with a bold font.

**Thank You!**

Your good work on this assignment helps us validate modifications of references to improve Wikipedia's quality. Thank you for your help!

**Figure A1 -** Instruction for crowdworkers from FigureEight Task (Part 1)

## Examples

Before starting, here are some examples of how to decide in different cases. Please make sure you understand all of them before starting the task.

**Example 1:**

http://www.reocities.com/wlorac/russbois.txt

book | url = http://www.reocities.com/wlorac/russbois.txt | title = a most noble anchorage : a story of russell and the bay of islands | last = king | first = marie | publisher = northland historical publications society | date = 1992

*For further information, you can check the revision on Wikipedia.*

**Please check if the above references refer to the same resource.** (required)
- ● Same resource
- ○ Different resources
- ○ Not clear (Please, do not overuse this option)

These two references refer to the **same resource** because they share the same URL.

**Example 2:**

https://www.captechu.edu/academics/graduate-academics/cyber-information-security-ms/

web | url = https://watson.okstate.edu/msia/ | title = m.s. in information assurance ( msia ) - oklahoma state university | website = watson.okstate.edu

*For further information, you can check the revision on Wikipedia.*

**Please check if the above references refer to the same resource.** (required)
- ○ Same resource
- ● Different resources
- ○ Not clear (Please, do not overuse this option)

These two references refer to **different resources** because they do not share the same URL and the first reference does not have a title or author name allowing any other judgment.

**Figure A2 -** Instruction for crowdworkers from FigureEight Task (Part 2).

**Example 3:**

These two references refer to the **same resource** because they share the same pmid.

**Example 4:**

These two references refer to **different resources** because they:

1) do not share the same URL,
2) do not share the same title.

In other words, they refer to 2 different documents: the Report from 2014 and the Report from 2016. You can open the URL-addresses (see below) and check that those are different documents. http://supremecourt.gov.bd/resources/contents/Annual_Report_2014.pdf and http://supremecourt.gov.bd/resources/contents/Annual_Report_2016.pdf

**Figure A3 -** Instruction for crowdworkers from FigureEight Task (Part 3).

**Example 5:**

These two references refer to the **same resource** because they:

1) share the same authors' names,
2) share the same title.

**Example 6:**

These two references refer to the **same resource** because they:

1) share the same title (nuance: the subtitle was added after the colon symbol in the second reference),
2) share the same URL.

**Figure A4 -** Instruction for crowdworkers from FigureEight Task (Part 4).

**Figure A5** - Example question of the task on FigureEight.

**Appendix B - Stratified sampling of references**

---

1. Define 8 buckets of references of size 125, one bucket per stratum.
2. Select random article A
3. Select random revision R in A
4. Select random reference f in R
5. Initialize reference S = R
6. Select the next revision S = S + 1:
   a. Select candidate references in revision S. A reference is a candidate c if there is no exact match in R.
      i. For each reference c calculate similarity between f and c
      ii. If the bucket (n=125) for similarity is not full, store the pair f and c.
   b. If no candidates are present in revision S then **go to step 6.**
7. If there are empty buckets left. **Go to Step 2.**
8. Otherwise (if all buckets are full), **Finish**

---

**Figure B1 - Stratified sampling of 1000 pairs of references.** The figure displays the pseudocode used for the sampling of references used for the evaluation of our method.

**Appendix C - Dataset of references with document identifiers.**

Besides our main dataset, that contains all changes of the references despite the presence or absence of document identifiers (DIDs), we also created the dataset where references between revisions were assumed to be *Equivalent* if they share the DID. The dataset included 7% of all the references. In this dataset, a reference would be created only when a DID is introduced. The dataset contains the following actions: 3,943,984 creations, 10,425,320 modifications, 2,685,187 deletions, and 1,566,989 reinsertions.

After matching these reference sequences with our full dataset we found that 10% of DID-References (Figure C1) did not have a DID in their first revision. Thus, references' creation date and time were wrong in 10% cases. All the actions (creation, modifications, deletions, reinsertions) that happened with this 10% of DID-References before their DID introduction would some up to 12.1% of actions of DID-References.

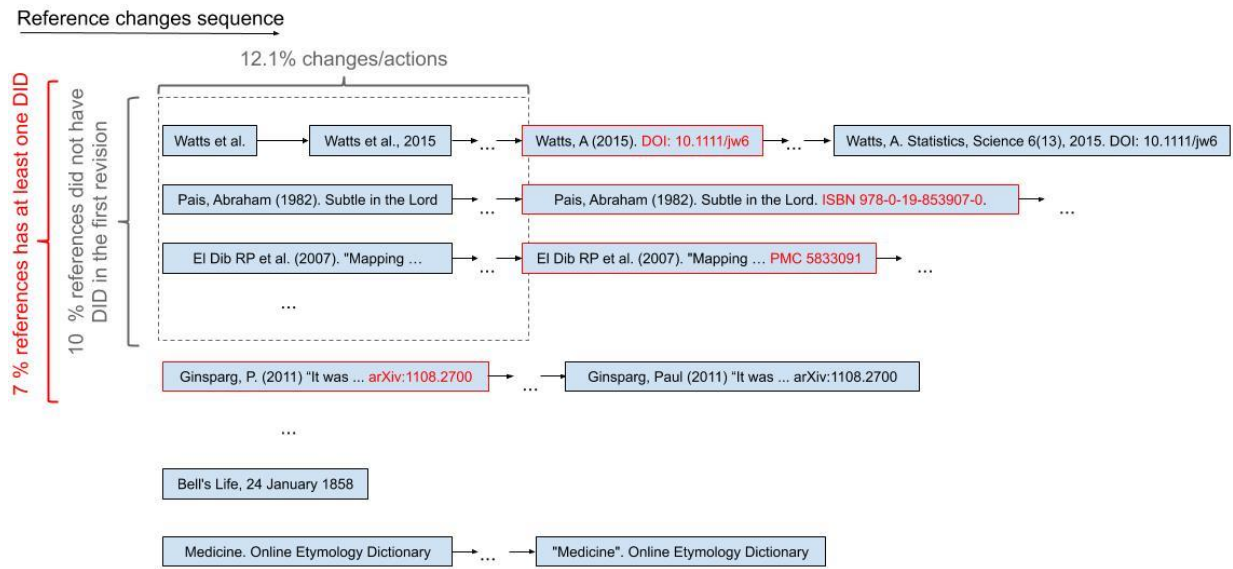

**Figure C1 - References change sequence examples**.

Initially, that dataset is supposed to solve the validation problem while document identifiers are the unique attributes of referenced objects. Performance indicators for actions are presented in Table C1, F1-score exceeds 0.97, 0.88 and 0.92 for insertions, deletions, and modifications accordingly. Nevertheless, one has to keep in mind that these indicators were calculated based on 7% of references. Thus, metrics might be biased towards references with (1) shorter strings, i.e. those with fewer details and attributes, (2) non-scientific objects.

**Table C1 -** Micro performance metrics for actions, where references with document identifiers were used as the gold standard dataset.

| | Reinsertion (n=1,566,989) | Deletion (n=2,685,187) | Modification (n=10,425,320) |
|---|---|---|---|
| **F1** | 0.97 | 0.88 | 0.92 |
| **Precision** | 0.97 | 0.88 | 0.93 |
| **Recall** | 0.97 | 0.87 | 0.91 |

**Appendix D - DIDs introduction time span**

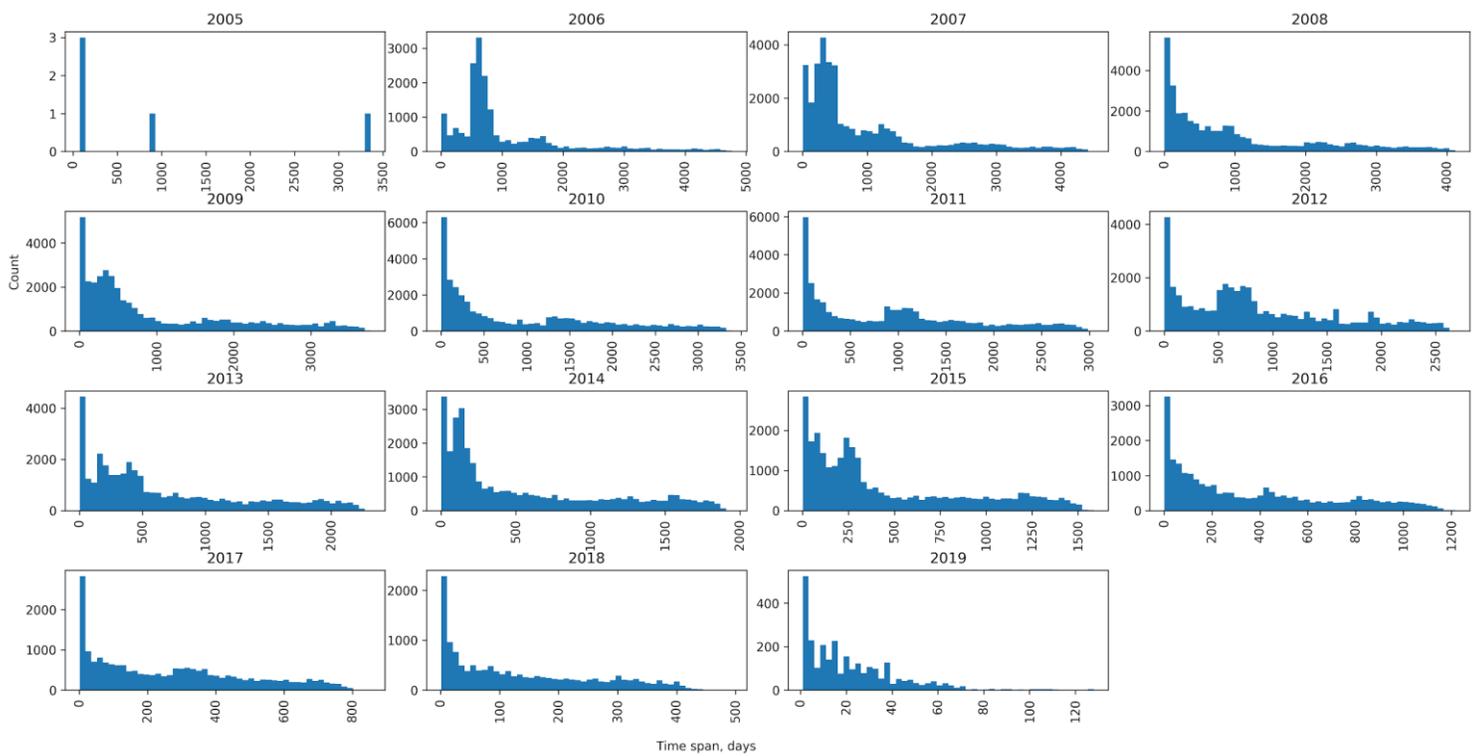

**Figure D1** - **DIDs introduction time span for 2005 - 2019 years of reference creation**. X-axis: the time difference between reference creation and DID introduction.

Zagovora et al., 2020
**Appendix E - The biggest peaks of added DIDs**

**Table E1 -** Represents months with the highest number of modifications when DIDs were added to existing references. The second column represents the number of references where DIDs were added. The third column represents which percentage of omitted DID-Rs remain omitted DID-Rs that month due to modifications that happened. For example, after modifications in May 2008, 81% of omitted DID-Rs left undiscoverable (meaning that they should have had DID but it had not been added yet).

| Date | Count | Remaining omitted DID-Rs, % |
|---|---|---|
| 2008-05 | 10037 | 81 |
| 2008-06 | 12091 | 74 |
| 2014-05 | 18131 | 83 |
| 2014-08 | 8101 | 91 |
| 2016-01 | 8451 | 89 |
| 2018-02 | 8662 | 83 |
| 2019-02 | 12486 | 44 |
| 2019-03 | 9797 | 11 |

**Appendix F -  List of bot names sources**

Not all bot accounts have a bot flag (so they can be listed in the list https://en.wikipedia.org/wiki/Special:ListUsers/bot ). So we created the list that would include all former and current bot names. This final list is a union of all the following sources (as of 01.08.2019):

1. https://stats.wikimedia.org/EN/TablesWikipediaEN.htm#bots
2. https://en.wikipedia.org/wiki/Wikipedia:List_of_bots_by_number_of_edits
3. https://en.wikipedia.org/wiki/Wikipedia:Bots/Status/inactive_bots_1
4. https://en.wikipedia.org/wiki/Wikipedia:Bots/Status/inactive_bots_2
5. https://en.wikipedia.org/wiki/Wikipedia:List_of_Wikipedians_by_number_of_edits/Unflagged_bots
6. https://en.wikipedia.org/w/api.php?action=query&list=allusers&augroup=bot
7. https://en.wikipedia.org/wiki/Category:Approved_Wikipedia_bot_requests_for_approval
8. https://en.wikipedia.org/w/api.php?action=query&list=categorymembers&cmtitle=Category:Approved_Wikipedia_bot_requests_for_approval&cmlimit=5000
9. https://en.wikipedia.org/wiki/Wikipedia:Bots/Requests_for_approval/Approved
10. https://en.wikipedia.org/wiki/Wikipedia:Bots/Status
11. https://en.wikipedia.org/w/index.php?title=Wikipedia:List_of_bots_by_number_of_edits/latest&oldid=185748540
12. https://en.wikipedia.org/w/index.php?title=Wikipedia:List_of_bots_by_number_of_edits&oldid=359820313
13. https://en.wikipedia.org/w/index.php?title=Wikipedia:List_of_bots_by_number_of_edits&oldid=271877315
14. https://en.wikipedia.org/wiki/Wikipedia:Bots/Requests_for_approval
15. https://en.wikipedia.org/wiki/Wikipedia:Bots/Requests_for_approval/Approved/Archive_14
16. https://en.wikipedia.org/wiki/Wikipedia:Bots/Requests_for_approval/Approved/Archive_13
17. https://en.wikipedia.org/wiki/Wikipedia:Bots/Requests_for_approval/Approved/Archive_12
18. https://en.wikipedia.org/wiki/Wikipedia:Bots/Requests_for_approval/Approved/Archive_11